\documentclass[usenatbib]{mn2e}
\usepackage{graphicx,color,epsfig}

\newcommand{\apj}{ApJ}
\newcommand{\apjs}{ApJS}
\newcommand{\mnras}{MNRAS}
\newcommand{\icarus}{ICARUS}
\newcommand{\aap}{A\&A}
\newcommand{\araa}{ARA\&A}
\newcommand{\apjl}{ApJL}

\newcommand{\nat}{Nature}

\newcommand{\planss}{Planetary and Space Science}

\def\ltsima{$\; \buildrel < \over \sim \;$}
\def\simlt{\lower.5ex\hbox{\ltsima}}
\def\gtsima{$\; \buildrel > \over \sim \;$}
\def\simgt{\lower.5ex\hbox{\gtsima}}

\def\msun{{\,{\rm M}_\odot}}

\newcommand\mearth{{\,{\rm M}_{\oplus}}}

\newcommand\mj{{\,{\rm M}_{\rm J}}}

\def\del#1{{}}

\title[Tidal Downising model I]{Tidal Downsizing model. I. Numerical methods:
  saving giant planets from tidal disruptions.}

\author[S. Nayakshin]{Sergei Nayakshin\\ 
Department of Physics \& Astronomy,
  University of Leicester, Leicester, LE1 7RH, UK\\
{E-mail:~} {\rm Sergei.Nayakshin@le.ac.uk}}

\begin{document}

\date{Received}

\pagerange{\pageref{firstpage}--\pageref{lastpage}} \pubyear{2008}

\maketitle

\label{firstpage}

\begin{abstract}
Tidal Downsizing (TD) is a recently developed planet formation theory that
supplements the classical Gravitational disc Instability (GI) model with
planet migration inward and tidal disruptions of GI fragments in the inner
regions of the disc. Numerical methods for a detailed population synthesis of
TD planets are presented here. As an example application, the conditions under
which GI fragments collapse faster than they migrate into the inner $a\sim$~a
few AU are considered. It is found that most gas fragments are tidally or
thermally disrupted unless (a) their opacity is $\sim 3$ orders of magnitude
less than the interstellar dust opacity at metallicities typical of the
observed giant planets, or (b) the opacity is high but the fragments accrete
large dust grains (pebbles) from the disc. Case (a) models produce very low
mass solid cores ($M_{\rm core}\simlt 0.1\mearth$) and follow a negative
correlation of giant planet frequency with host star metallicity. In contrast,
case (b) models produce massive solid cores, correlate positively with host
metallicity and explain naturally while giant gas planets are over-abundant in
metals.
\end{abstract}


\section{Introduction}\label{sec:intro}

A most general description of a planet is that it is a self-gravitating object
composed of a heavy element core and an envelope of gas. Terrestrial like
planets are dominated by solid cores whereas giant gas planets are mainly
hydrogen gas. Given these observational facts, it should come as no surprise
that there are two competing scenarios for planet formation that take opposite
views on what comes first \citep[for a recent review see][]{HelledEtal13a}. In
the top-down scenario hydrogen gas cloud comes first and the solid element
core is assembled later. In the bottom-up picture the solid core must be made
first before a bound gas envelope appears. In the former scenario planets {\em
  loose} mass, especially gas, as they mature, whereas in the latter planets
gain mass with time.

The top-down hypothesis takes roots in the \cite{Kuiper51} proposition that
planets begin their existence as self-gravitating condensations of $\sim 3$
Jupiter masses of gas and dust formed in the Solar Nebula by Gravitational
Instability (GI) of the disc.  \cite{McCreaWilliams65} showed that microscopic
grains grow and sediment to the centre of such gas clumps within a few
thousand years, presumably forming a massive solid core there
\citep[cf. also][]{Boss97,Boss98}. These cores are the seeds of
terrestrial-like planets in the model.  \cite{Kuiper51} and
\cite{McCreaWilliams65} proposed that these cores could be all that is left of
the original gas protoplanets if the dominant gas component is disrupted by
tidal forces from the Sun \citep[this process was rediscovered
  by][]{BoleyEtal10}. It is natural in this picture that giant planets in the
Solar System are located further away from the Sun than terrestrial-like
planets.

On the other hand, \cite{Safronov69} instead posited that microscopic dust
grains in the protoplanetary disc grow larger and then somehow collect into
huge rocks of at least $\sim$ 1 km size, called planetesimals. These coalesce
into even larger solid cores. Low mass cores become terrestrial
planets. Largest cores, of mass $M\simgt 10 \mearth$ (10 Earth masses),
attract gaseous atmospheres from the protoplanetary disc and end up as giant
gas planets \citep[e.g.,][]{PollackEtal96}. This bottom-up scenario is now
called Core Accretion \citep[e.g.,][]{AlibertEtal05,MordasiniEtal09a} and is
by far the most popular planet formation theory.

Differentiation between these two theories was thought to be straight forward
based on the Solar System data. GI unstable discs were argued not to produce
self-gravitating clumps at all at the location of the Solar System planets due to an
inefficient disc cooling \citep{Gammie01,Rafikov05,Rice05}, so the sequence of
events envisioned by the top-down picture could not be even started. CA
picture, on the other hand, proved quite successful in explaining the Solar System
\citep{PollackEtal96,TsiganisEtal05}.

However, the above criticism of the top-down scenario neglects the possibility
of planet migration \citep[that is, shrinking of the planet's orbit due to
  gravitational torques from the protoplanetary disc,
  see][]{LinPap79,GoldreichTremaine80}. CA planets were equipped with
migration since \cite{Lin96}, who showed convincingly that hot jupiters could
not have formed at their observed planet-star separations, $a\sim 0.1$~AU, and
are more likely to have migrated from their assumed birth location of $\sim 5$~AU. 

In contrast, the role of migration for GI planet formation model was not
appreciated until \cite{BoleyEtal10,Nayakshin10c}. These authors pointed out
that gravitationally unstable discs do not actually need to form gas clumps at
$\sim$ a few AU to explain the observed planets there: in simulations, most GI
fragments are found to migrate rapidly from their birth locations at $\sim
100$~AU into the inner disc
\citep[e.g.,][]{VB06,BaruteauEtal11,ChaNayakshin11a,ZhuEtal12b}. It is hence
plausible that all of the observed giant planets were hatched by GI in the
outer disc and were then pulled much closer to the host star by the
gravitational disc torques. Furthermore, some of the fragments could give
birth to Earth-mass or more massive cores by grain growth and sedimentation,
and be tidally disrupted, potentially providing a "new" pathway\footnote{The
  author of this article, embarrassingly, did not know of the \cite{Kuiper51}
  and \cite{McCreaWilliams65} suggestions until they were pointed out to him
  by I. Williams after a seminar at the Queen Mary University in the fall of
  2010. TD is essentially the original \cite{Kuiper51} suggestion plus "super
  migration" of planets from $\sim 100$~AU to arbitrarily close to the star.}
to forming all kinds of planets at all separations in a single framework that
was called ``Tidal Downsizing'' (TD).

We note in passing that \cite{BowlerEtal14} recently presented the results of
the PALMS survey which shows that the frequency of giant gas planets at large
separations ($\sim 10-100$~AU) is very low, e.g., less than $\sim 10$\%,
implying that the ``poster child'' GI-planet system HR 8799
\citep{MaroisEtal08} is very rare. \cite{BowlerEtal14} conclude that ``disc
instability is not a common mode of giant planet formation''. In the context
of TD hypothesis, the observations of \cite{BowlerEtal14}, unfortunately, do
not tell us directly about disc fragmentation properties at these separations;
they rather show that GI-planets rarely survive at their initial large
separations to the present day. In fact, given that the inward migration times
of GI planets are as short as $\sim 0.01$ Million years
\citep{BaruteauEtal11}, it has been argued that it is not clear how any of the
observed GI planets (such as the multi-planetary system HR 8799) survive. For
example, \cite{ZhuEtal12b} found that all of their 13 simulated gas clumps
were either tidally destroyed, migrated into the inner few AU, or became brown
dwarfs due to gas accretion. Observations of \cite{BowlerEtal14} are therefore
in excellent agreement with expectations of TD. Any GI model that does not
include migration of planets should be treated with suspicion at this day and
age when an extremely rapid migration of giant planets in self-gravitating
accretion discs has been confirmed by around a dozen independent research
teams \citep[to add to the references above, see
  also][]{MayerEtal04,MichaelEtal11,MachidaEtal10,NayakshinCha13,TsukamotoEtal14}.

The potential of the top-down scenario remains poorly explored to this day,
mainly because understanding of different aspects of the model is not yet
complete, and connecting them together in order to make solid observational
predictions is not trivial.  The first population synthesis model for TD
hypothesis was presented by \cite{ForganRice13b}, who used a semi-analytical
approach to disc evolution and an analytical description for fragment
contraction, grain growth and the resultant core formation. The results of
their study are not particularly encouraging for TD. Many fragments were found
to run away in mass by accreting gas and became brown dwarfs at large
separations from the parent star. Others were disrupted by tides before they
could assemble a massive core. While solid cores with masses up to $\sim
10\mearth$ were assembled inside the fragments by grain sedimentation, most of
them were locked inside brown dwarf mass fragments. Out of a million gas
fragments simulated, only one yielded an Earth mass core {\it without} an
overlaying massive gas envelope. More recently, population synthesis was also
performed by \cite{GalvagniMayer14}, who used different prescriptions for
fragment cooling and did not address core formation. In contrast to
\cite{ForganRice13b}, the latter study found that most of the gas fragments
yield hot jupiters rather than brown dwarfs at large separations.

This large difference in the results of the two studies motivates the present
paper, which improves on both by using 1D numerical rather than analytical
descriptions of the coupled disc and planet evolution. The eventual goal is to
have a physically constrained and therefore predictive model of the TD
hypothesis, put in a computationally expedient framework, and enabling
statistical comparisons to the observed planetary data.

While a detailed comparison to the data is to be reported in follow up
publications, as a way of illustrating the methods and the information
obtained at the end of the runs, the question of GI fragments surviving 
the initial rapid inward migration from their birth place into the inner 
disc regions is considered. More specifically, the focus is on how and when 
the fragments can contract and collapse more rapidly than they migrate in.

This topic has direct observational connections.  Giant planets are observed to be
much more frequent around metal-rich stars than around metal-poor ones
\citep{Gonzalez99,FischerValenti05}. This correlation has been argued to
provide a direct support to CA theory since metal-rich environments assemble
massive cores much more readily
\citep[e.g.,][]{IdaLin04,IdaLin08,MordasiniEtal09a}. At the same time,
Gravitational Instability planets (and thus TD planets also) are expected to
survive the inward migration phase most readily in metal-poor environments
because their radiative cooling is the fastest at low dust opacities
\citep{HB11}. This therefore appears to contradict the observed
planet-metallicity correlation and is a very clear challenge to the TD
hypothesis.

Accretion of ``pebbles'', $\sim 0.1 - 10$~cm sized grains, from the
surrounding protoplanetary disc onto solid bodies, e.g., large planetesimals
up to massive cores, has been studied by a number of authors
\citep[e.g.,][]{JohansenLacerda10,OrmelKlahr10,LambrechtsJ12,LJ14,JohansenEtal14a}. The
process is interesting for many reasons, including explaining a potentially
more rapid growth of massive cores at $\sim$ tens of AU than by accretion of
planetesimals \citep{JohansenEtal14a} and may contribute to formation of
Uranus and Neptune \citep{HB14}.

\cite{Nayakshin14c} argued that while pre-collapse giant planets may be
inefficient gas accretors (see \S \ref{sec:broader} further on this), they
nevertheless should accrete pebbles for same reasons as solid cores in the CA
picture do, except at higher rates because TD fragments have much higher
($\sim 1$ Jupiter) masses. Adding a pebble injection term to the equations of
an otherwise isolated TD fragment evolution, a surprising result was obtained:
the fragments actually accelerated their contraction and collapsed sooner, not
later, despite the increase in the dust opacity associated with the higher
metallicity of the fragments. \cite{Nayakshin14c} explained the effect in
terms of the extra weight and (negative, of course) potential energy that
pebbles bring with them into the gas fragments. Pre-collapse fragments turn
out to be very sensitive to pebble accretion because they are nearly polytropes
with low effective adiabatic indexes, $\gamma\approx 1.4$. It was found that
addition of $\sim 10$\% by mass in grains is sufficient to drive a
typical pre-collapse molecular H fragment into collapse.

Even more recently, \cite{Nayakshin14d} presented a population synthesis like
study that included the ``metal overload'' collapse due to pebble accretion on
TD planets for the first time, and found a strong positive rather than
negative correlation of giant planet survival with disc metallicity. His model
did not include the processes of grain growth, sedimentation and core
formation inside the fragment. 

This paper expands the work of \cite{Nayakshin14d} by adding core growth
processes. Further, two ways of enhancing the survived giant planet fraction are considered: pebble accretion and
lower dust opacity due to grain growth \citep[this was not considered in][at all]{Nayakshin14d}. It is argued that low opacity models require
unrealistically small dust opacities ($\sim 3$ orders of magnitude smaller
than the interstellar grain opacity at same metallicity) to save the gas
fragments from disruptions. Even if this were plausible, low opacity models
suffer from a host of other issues: (i) predict a negative correlation of
planet frequency with host metallicity, (ii) cannot produce solid cores more
massive than $M_{\rm core}\sim 0.1\mearth$, (iii) fail to explain why the
observed giant planets are strongly over-abundant in metals. It is thus
clear that low opacity models are a dead end as far as an all inclusive TD
model for planet formation is concerned.

Pebble accretion models, on the other hand, offer a more attractive
alternative. They do not require unrealistically small opacities, naturally
explain why giant planets can be much more metal rich than their host stars,
and produce a positive giant planet-metallicity correlation. Even the
``discarded'' gas fragments, e.g., those destroyed by tides or thermal
over-heating are important since the masses of solid cores surviving such
disruptions can be in the terrestrial and above mass range.

The paper is structured as following. A brief introduction into the physical
motivation behind the TD hypothesis, and a short overview of the literature on
TD, including potential applications to observations of the Solar System and
beyond, is given in \S 2 below. Numerical algorithms for simulating the
planet-disc interaction and the disc evolution are introduced and some example
(point mass planet) migration tracks are calculated in \S \ref{sec:disc}. \S
\ref{sec:planet} deals with the methods used to calculate the fragment
evolution, including pebble accretion, grain growth, settling and core
formation, whereas \S \ref{sec:together} explains how the two modules are put
in one code and what the initial conditions are. \S\S \ref{sec:opacity} and
\ref{sec:pebbles} show examples of fragment evolution tracks for low opacity
and pebble accretion models, respectively. In \S \ref{sec:discussion}, the
implications of the results obtained for TD hypothesis is given.

\section{Physical motivation}\label{sec:big_idea}

\begin{figure}
\centerline{\psfig{file=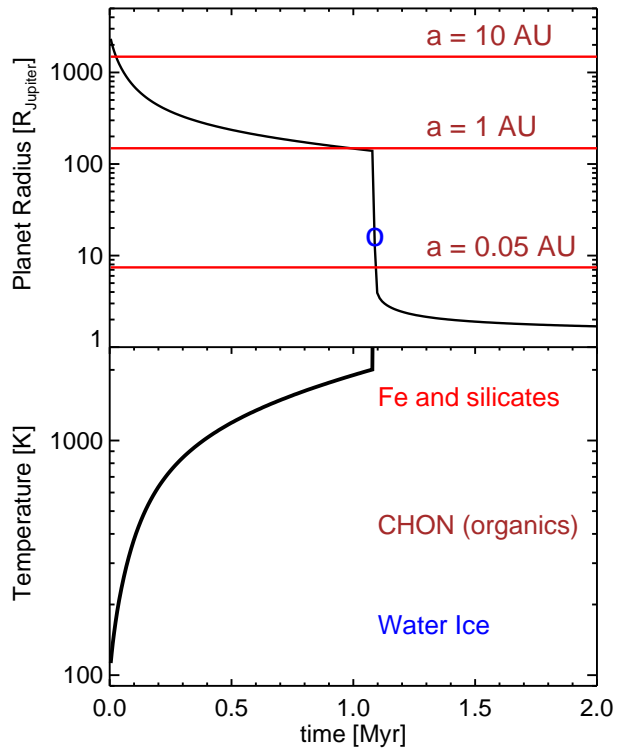,width=0.5\textwidth,angle=0}}
\caption{Radiative cooling of an isolated Jupiter mass gas giant planet. The
  planet is coreless and of Solar composition. The upper panel shows planet's
  radius (black curve). The three horizontal lines depict the planet's Hills
  radius if the planet were in orbit around a $1 \msun$ mass star at distances
  indicated just above the lines. The lower panel shows evolution of the
  central temperature of the planet. During the first Million years, the
  planet is dominated by molecular hydrogen. The three different grain species
  considered here condense out below temperatures indicated approximately by
  the location of the species name in the panel.}
\label{fig:1}
\end{figure}

\subsection{Isolated planet evolution}\label{sec:isolated}

Figure \ref{fig:1} shows the evolution of an isolated Jupiter mass planet
contracting from an initial state in which its central temperature is 100
K. The planet's metallicity is set to Solar, and opacity is set to that of the
interstellar gas/dust mix \citep{ZhuEtal09}.  The calculation is done with the
part of the code described in section \ref{sec:planet}, although grain growth
and grain sedimentation within the planet are neglected. The top panel of
figure \ref{fig:1} shows the evolution of planet's radius, $R_p$, in units of
Jupiter's radius, $R_J$, whereas the bottom panel shows the planet's central
temperature.

The planet's evolution is conveniently described in two stages
\citep{Bodenheimer74}. In the first "pre-collapse" stage, the planet's central
temperature is below $\sim 2000$~K, and Hydrogen is in the molecular form. As
temperature exceeds $\sim 2000$~K, H$_2$ molecules begin to dissociate. The
process of breaking the molecules up requires 4.5 eV per Hydrogen molecule,
which is very large: it is equivalent to the thermal kinetic energy of two H
atoms at temperature $T\approx 10^4$ K. The dissociation process is therefore
a huge energy sink, and leads to the planet loosing its hydrostatic
equilibrium and collapsing dynamically. The new, post-collapse, equilibrium is
such that H is atomic and partially ionised, central planet's temperature is
well above $10^4$~K, and the planet radius is more than ten times smaller than
in the pre-collapse state. In Fig. \ref{fig:1}, the dynamical collapse
corresponds to the nearly vertical part of the radius and temperature
evolution curves. After the collapse, the planet arrives on the post-collapse
stage, which is one and the same as the "hottest start" track for giant gas
planets \citep[e.g.,][]{MarleyEtal07}. In Fig. \ref{fig:1}, this post-collapse
part of the evolution begins at the blue "o" symbol in the top panel of the
figure. During this phase, the planet contracts from $R_p \simgt 10 R_J$ to
$R_p\sim 2 R_J$ relatively rapidly, e.g., in some $10^5$~years, and then takes
much longer to eventually contract to $1 R_J$.

\subsection{Tidal disruption of protoplanets}\label{sec:destroy}

The key point to observe from the top panel of figure \ref{fig:1} is just how
extended the planet is at birth ($R_p \sim$ 1 AU, or about $2000 R_J$), and
that it takes over a Million years for the planet to contract to radii
comparable to that of Jupiter. The red solid horizontal lines show the Hill's
radius, $R_H$, of the planet at three distances from a $M_*= 1\msun$ host
star. When $R_p$ is larger\footnote{this condition may be even stricter for
  young rotating planets, e.g., $R_p\simgt 0.5 R_H$ may be sufficient for
  disruption, see \S \ref{sec:tides}  below} when $R_H$, tidal forces from the star exceed
self-gravity of the planet, and it can be disrupted, as in the original
\cite{Kuiper51} scenario. Figure \ref{fig:1} thus shows that the planet can be
tidally disrupted at the planet-star separations $1 \simlt a \simlt 10$~AU in the
pre-collapse stage. In addition, the planet can be tidally disrupted in the
post-collapse stage in the innermost "hot" region of the disc, $a\simlt
0.1$~AU, provided the collapse happened very recently (no more than $\sim
10^5$~years earlier). \cite{Nayakshin11a,NayakshinLodato12} proposed that such hot
disruptions may be the origin of "hot" sub-giant planets such as hot
Super-Earths or hot Neptunes.

Another point worth noting from figure \ref{fig:1} is that the astrophysical
metal components of the protoplanet (that is, all elements heavier than H and He)
can settle into the centre of it provided that condensation temperature of the
components is higher than the central temperature of the planet. In this paper
we limit our attention to three dominant grain species: water ice, organics
called CHON, and a mix of Fe and silicates (cf. \S \ref{sec:gr_growth}
below). The condensation temperatures for these species are marked
approximately by the position of the respective text in the bottom panel of
figure \ref{fig:1}. Note that the planet spends little time in the cold
configuration, since it cools relatively rapidly initially. From this one can
expect water to be the least able to condense down in TD planets, whereas Fe
and silicates be the most able to do so \citep[this agrees with results of][that TD cores are mainly composed of rocks ]{ForganRice13b}.

Grain sedimentation may lead to core formation inside of the protoplanet
\citep[e.g.,][]{McCreaWilliams65,Boss97,HS08}. If the gaseous component of the
protoplanet is then disrupted by the tides, the nearly ``naked'' core
survives, since its density is much higher. Such disruptions may be an
alternative to CA origin for terrestrial-like planets
\citep{Kuiper51,BoleyEtal10,Nayakshin10c}.

\subsection{TD model possible outcomes}\label{sec:simple}

It is clear that the fate of a gas fragment formed in the outer disc by GI is
sealed \citep{BoleyEtal10,ForganRice13b} by the ratios of the various time
scales: $t_{\rm coll}$, the planet's contraction and collapse time scale;
$t_{\rm migr}$, the planet's migration time scale; and $t_{\rm sed}$, the
grain growth and sedimentation time scale.  Qualitatively, there are several
possibilities:
\begin{enumerate}
\item If the migration time is the shortest of the three, the fragment is
  disrupted with essentially nothing remaining of it;
\item if $t_{\rm sed} < t_{\rm migr} < t_{\rm coll}$, then grains sediment
  down before the planet is disrupted. There is therefore a remnant, a solid
  core \citep{BoleyEtal10}, possibly surrounded by a post-disruption
  atmosphere remaining bound to the core if the latter is massive enough;
\item If $t_{\rm coll} < t_{\rm migr}$, then the planet collapses before it is
  disrupted, and survives as as a gas giant planet, provided it is not
  disrupted in the "hot" region or pushed all the way into the star.
\item If grain sedimentation time is the shortest of the three, and the core
  mass is significant, $M_{\rm core}\sim $~a few to ten or more Earth masses,
  then the grain component can affect the whole fragment by either triggering
  its collapse \citep{NayakshinEtal14a} or destruction \citep{NayakshinCha12}.
\end{enumerate}

Clearly, it is essential to be able to calculate not only these three time
scales but also the detailed planetary structure as well as the protoplanetary
disc evolution. The planet formation/destruction road map outlined above is only a
very rough guess as to what may actually happen in a realistic disc and planet
setting. 

\subsection{Broader connections of TD}\label{sec:broader}

\cite{VB05,VB06} found formation and a rapid inward migration of massive
self-gravitating gas clumps in their 2D simulations of protostellar discs. They
argued that clump destruction episodes could be related to the FU Ori
outbursts of young stars.  \cite{BoleyEtal10} were first to point out that gas
clump migration, core formation and then envelope disruption may result in
formation of terrestrial-like planets. \cite{BoleyDurisen10} noted that
dynamics of solids can be important enough to change the fragmentation
properties of the gravitationally unstable discs, and to affect the gas clump
compositions. These in general do not therefore have to be same as that of
their parent discs \citep{BoleyEtal11a}.

\cite{NayakshinEtal11a,Vorobyov11a,BridgesEtal12a} argued that the
high-temperature environment inside the clumps (before they are disrupted) may
provide a natural nursery for high temperature inclusions found in chondrules
of the Solar System. \cite{Nayakshin11a} proposed that prograde rotation of
the pre-collapse gas fragments may have been imprinted on the spin directions
of Earth and Mars, and yield enough angular momentum for proto-Earth to revive
the fission hypothesis for the Moon formation.

\cite{MachidaEtal11} observed recurrent episodes of clump formation,
destruction and protostellar outflow injections in 3D simulations of
self-gravitating discs. \cite{BaruteauEtal11,MichaelEtal11} made detailed
studies of planet migration in gravitationally unstable discs, finding
migration times of order $\sim 10$ thousand years. \cite{ChaNayakshin11a}
simulated in 3D a gravitationally unstable gas disc together with the grains
that were allowed to grow, and found that grains grew most rapidly inside the
clumps and indeed sedimented into the dense cores. \cite{GalvagniEtal12} made
first 3D simulations of collapse of the rotating molecular clumps, noting the
importance of the angular momentum and formation of the post collapse
circum-planetary discs.

It is important to note that planetesimals are not a pre-requisite for TD
model, unlike for CA. In the context of TD, minor solids, such as asteroids,
are remnants of larger solid bodies that were fragmented in collisions. Thus,
\cite{NayakshinCha12} argued that Kuiper and asteroid belts in the Solar
System could be made in disruptions of molecular gas fragments in which some
of the solids sedimented into the centre and got locked into a number of Ceres
mass bodies. They found that after the gas clump disruption, the distribution
of solid bodies naturally forms rings with sharp edges, as observed in the
Solar System, and that this model does not have the ``mass budget deficit'',
resolving the two well known problem of the classical models of the Kuiper
belt \citep[e.g.,][]{Morbidelli08}.


\cite{NayakshinCha13} argued that gas accretion on relatively low mass gas TD
fragments ($M_p \simlt$~a few $\mj$) is inefficient due to preheating effects
from the planet on surrounding gas. These authors found that there is a
dichotomy in the fate of the clumps. Low mass clumps were found to migrate in
rapidly at more or less constant mass, and eventually were destroyed by the
host star's tides. Clumps more massive than $\sim 5 \mj$ were found to run
away in the ``opposite'' sense: accreting gas rapidly, they become as massive
as $\sim 50 \mj$, open deep gaps in their protoplanetary discs, and stall at
about the initial star-clump separation. This divergent evolution may explain
the divergent results on the fate of the gas clumps in the literature
\citep[giant planets or brown dwarfs -- see][for references]{NayakshinCha13},
since small changes to the disc parameters and the treatment of gas
thermodynamics may shift the clumps from one of the other regime.

\cite{TsukamotoEtal14} presented 3D radiative hydrodynamics simulations of
massive self-gravitating discs. Their approach is clearly preferable to all of the
previous 2D studies \citep[e.g.,][]{ZhuEtal12b} and also the earlier 3D studies which used
radiative cooling prescriptions \citep[e.g.,][]{ChaNayakshin11a}. The RHD
simulations could in principle directly address many of the issues needed for
a quantitative TD population synthesis models, such as the initial clump mass
\citep[e.g.,]{ForganRice11,ForganRice13}, and the migration and clump
accretion history. \cite{TsukamotoEtal14} find that most of their clumps are
as massive as $\sim 3\mj$ at birth and grow rapidly in mass while migrating
in. What is not clear is how general these results are with respect to changes
in initial conditions, and how they depend on the dust opacity model assumed
by \cite{TsukamotoEtal14}.

Here, and everywhere in the paper, only the simplest version of events is
considered, in which the planets start out as self-gravitating gas fragments
and do {\it not accrete} gas or planetesimals from the disc. Both gas
accretion and especially planetesimals are virtues of the Core Accretion model
\citep[e.g.,][]{PollackEtal96}. It appears to be most prudent to first
consider the simplest ``Orthodox'' TD hypothesis. More complicated
versions of TD may be considered in the future.

\section{Disc evolution -- planet migration module}\label{sec:disc}

This section presents numerical methods with which planet migration and disc
evolution are modelled in the paper. As far as the planet-disc interaction is
concerned, the planet is treated as a point mass. The evolution of the
internal planet's variables is considered in \S \ref{sec:planet}.

Using a 1D viscous disc evolution approach, \cite{NayakshinLodato12} studied
planet migration and planet mass loss for very massive ($M_p\sim 10 M_J$) gas
planets in the "hot" disc region, e.g., at $a\sim 0.1$~AU. Such massive gas
fragments are always in the type II migration regime, when they open a deep
gap in the disc \citep{LinPap86}. \cite{NayakshinLodato12} were interested not
only in the fate of the planets but also in the potential connection between
giant planet disruptions and the FU Ori outbursts of young stars
\citep{VB06,BoleyEtal10,Nayakshin11b}. Therefore, the planet-disc mass
exchange was followed in detail and on short time scales (much shorter than a
year) to compare them with the observed light curves and spectra of some
observed FU Ori sources \citep[e.g.,][]{CLEtal07,EisnerH11}.

The approach here follows that of \cite{NayakshinLodato12} with a few
changes. Since the fate of less massive gas planets and also that of even less
massive remnants (solid cores, if the planets are disrupted) is of interest to
us here, type I migration regime (no gap in the disc) is also
included. Accretion luminosities of young stars, although calculated here
automatically by the viscous disc evolution code, are of no direct
interest. Therefore, the treatment of the planet's mass loss is simplified
from that of \cite{NayakshinLodato12}: it is assumed that the planet is
disrupted instantaneously if the disruption criteria are satisfied, and the
mass deposition back in the disc is neglected for simplicity.

The protoplanetary disc is described by a viscous azimuthally symmetric one
dimensional time-dependent model that encapsulates the standard accretion disc
solution of \cite{Shakura73} with addition of the tidal torque of the planet
on the disc. The disc surface density, $\Sigma(R)$, is evolved in this
approach according to
\begin{equation}
  \frac{\partial\Sigma}{\partial
  t}  = \frac{3}{R}\frac{\partial}{\partial R}
  \left[R^{1/2}\frac{\partial}{\partial R}(R^{1/2}\nu\Sigma)\right]
  -\frac{1}{R}\frac{\partial}{\partial
  R}\left(2\Omega R^2\lambda\Sigma\right) 
\label{eq:diffplanet}
\end{equation}
where $\Omega(R) = \sqrt{GM_*/R^3}$ is the Keplerian angular frequency at
radius $R$, viscosity $\nu = \alpha_{SS} c_s H$, where $c_s$ and $H$ are the
midplane sound speed and the disc height scale, $\lambda=\Lambda/(\Omega
R)^2$, and $\Lambda$ is the specific tidal torque from the planet. The latter
is a weighted sum of the type I and the type II contributions,
\begin{equation}
\lambda = \lambda_I \left(1-f_{II}\right) + \lambda_{II} f_{II}\;,
\end{equation}
where $0\le f_{II}\le 1$ is a switch controlling whether the planet migrates
in type II ($f_{II} =1$) or type I ($f_{II}=0$) regimes. Two-dimensional
simulations \citep{CridaEtal06} show that a deep gap in the disc is opened
when parameter 
\begin{equation}
{\cal P} = {3H\over 4R_H} + 50 \alpha_{SS} \left({H \over a}\right)^2 {M_*
  \over M_p} \simlt 1\;,
\label{pgap}
\end{equation}
 where $H$ is the disc vertical scaleheight at planet's location, $a$, and
$\alpha_{ss}< 1$ is the Shakura-Sunyaev viscosity parameter. We therefore set
\begin{equation}
f_{II} = \min\left\{1, \exp\left[ - \left({\cal P}-1\right)\right]\right\}\;.
\label{kill2}
\end{equation}
The normalised specific torque for type II migration is given by the widely
used expression,
\begin{eqnarray}
\label{eq:torque}
\lambda_{II}= & \displaystyle\frac{q^2}{2}\left(\displaystyle
\frac{a}{\Delta R}\right)^4  &
R>a \\
\nonumber \lambda_{II}= & -\displaystyle\frac{q^2}{2}\left(\displaystyle 
\frac{R}{\Delta R}\right)^4  & R<a. 
\end{eqnarray}
\citep{armibonnell02,LodatoClarke04,AlexanderEtal06}, where $\Delta R =
R-a$. We smooth the torque term for $R\approx a$, where it would have a
singularity (see equation \ref{eq:torque}). We use the same smoothing
prescription as \citet{SyerClarke95} and \citet{LinPap86}, i.e. for
$|\Delta R| <\max[H,R_H]$, where $H$ is the disc thickness and $R_H = a(M_{\rm
  p}/3M_*)^{1/3}$ is the size of the Hill sphere (Roche lobe) of the planet.

As far as we are aware, there is no corresponding expression for
$\lambda_{I}$, perhaps partly because one usually studies type I migration in
the limit where the planet's mass is small compared to the disc mass, so that
the back reaction of the planet on the disc can be neglected. Our planets can
be massive and yet still be in type I regime, especially in the outer
self-gravitating disc \citep{BaruteauEtal11}, and also in the inner disc if
viscosity parameter $\alpha_{SS}$ is "large" (e.g., greater than $\sim$
0.01-0.03). Furthermore, gas giant planets studied here are bright and can
easily dominate the disc luminosity around their location, e.g, within the
region $|R-a|<\max[H,r_H]$ \citep[see][]{NayakshinCha13}. This may affect the
thermodynamic disc structure near the location of the planet, yet none of the
migration disc studies in the literature currently take this effect into
account.

Faced with this uncertainty, we chose to use the simplest approach that
appears reasonable and is to be improved in the future as details of type I
migration are understood further. The total torque between the disc and the
planet, $\Gamma_I$, in a self-gravitating outer disc \citep{BaruteauEtal11} is
similar to the standard isothermal disc result \citep{BateEtal03}, which
yields type I planet migration time scale ($t_I \equiv - a/(da/dt)_I$)
\begin{equation}
t_{I} = {M_*^2 \over M_p M_{d}} {a^2\over H^2} \Omega_a^{-1}\; \left(1 + {M_p
  \over M_d}\right)\;,
\label{time1}
\end{equation}
where $M_d = \pi \Sigma a^2$ is approximately the local disc mass, $\Omega_a$
is the local Keplerian angular velocity, and the $1+M_p/M_d$ factor is
introduced to account for a possibility that $M_p/M_d \gg 1$ for our planets,
which is not the typical case for type I migrating planets \citep[e.g., see \S
  2.1 in][]{BaruteauEtal14a}. While the form of the factor $1 + M_p/M_d$ is
debatable, we note that the main conclusions of this paper do not depend on
this factor at all since the migrating planets are always much less massive
than the disc while migrating most of their way in.

We assume that only the material close to the planet's location
exchanges angular momentum with it, and that the strength of the interaction
falls off away from the planet as
\begin{equation}
\lambda_I = \lambda_{I}' \exp\left[ - {|\Delta R|\over \Delta R_I}\right]\;,
\label{lambda1}
\end{equation}
where $\Delta R = R-a$, and $\Delta R_I = H + R_H$. $\lambda_I'$ is found by
requiring the total type I torque from the planet on the disc to equal that
(with the minus sign) from the disc on the planet, that is, $M_p (G M_*
a)^{1/2}/(2 t_I)$. In equation \ref{time1}, $\Omega_a = \sqrt{GM_*/ a^3}$ is
the local Keplerian angular velocity.

The migration rate of the planet is calculated from the conservation of the
total (disc plus planet) angular momentum as in \cite{NayakshinLodato12}.
Also as in \cite{NayakshinLodato12}, see their \S 3.2, the vertical structure
of the disc is calculated as in the standard \cite{Shakura73} model, taking
into account the irradiation from the star, which can dominate the disc
heating at larger radii. The disc thermal balance equation takes into account
the finite thermal disc time scale, $(\alpha_{SS} \Omega(R))^{-1}$, which
determines how quickly the disc achieves thermal equilibrium if accretion rate
varies suddenly. This is important in the inner disc, if the disc accretion
rate is high and the thermal disc instability is triggered \citep{Bell94}. The
disc opacity is from \cite{ZhuEtal09}, multiplied by the ratio of the disc
metallicity to the Solar metallicity, $z_\odot=0.015$.

Finally, it should be noted that in this paper only the outcome of the planet
rapid inward migration phase from the outer into the inner disc is studied,
and a longer time scale evolution of the survived fragments is not
addressed. This would require a better treatment of type I migration, which
currently remains controversial and model dependent especially for the more
massive planets \citep{BaruteauEtal14a}, and a disc photo-evaporation term to
be added to equation \ref{eq:diffplanet} to model late disc evolution. While
this is relatively straight forward as disc dispersal by photo-evaporation is
believed to be reasonably well understood \citep{AlexanderREtal14a}, we wish
to avoid introducing further parameters and complications to our model at this
initial stage.

\subsection{Example disc evolution and planet migration tracks}\label{sec:2discs}

\begin{figure*}
\begin{minipage}{3.2in}
\psfig{file=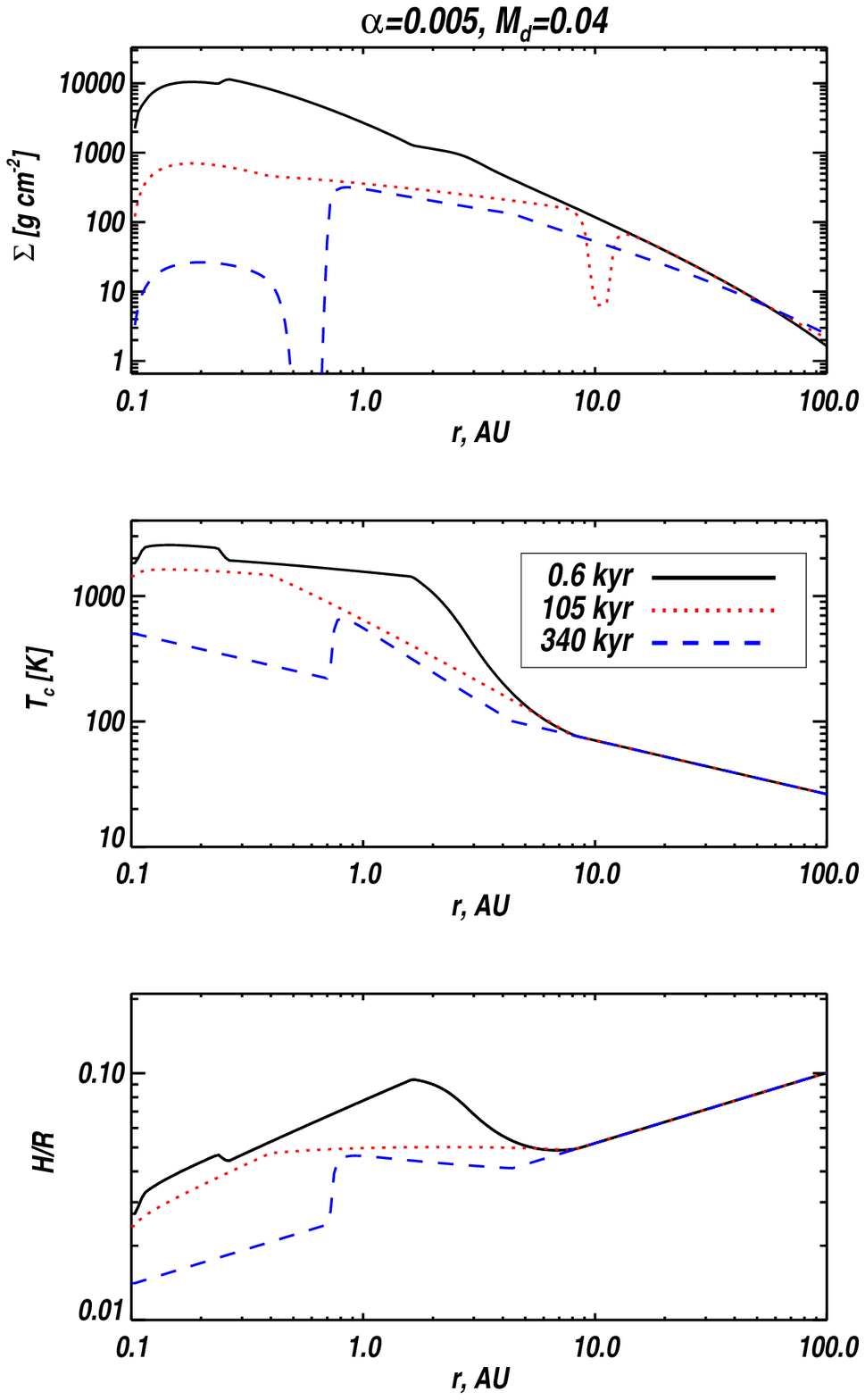,width=0.99\textwidth,angle=0}
\end{minipage}
\begin{minipage}{3.2in}
\psfig{file=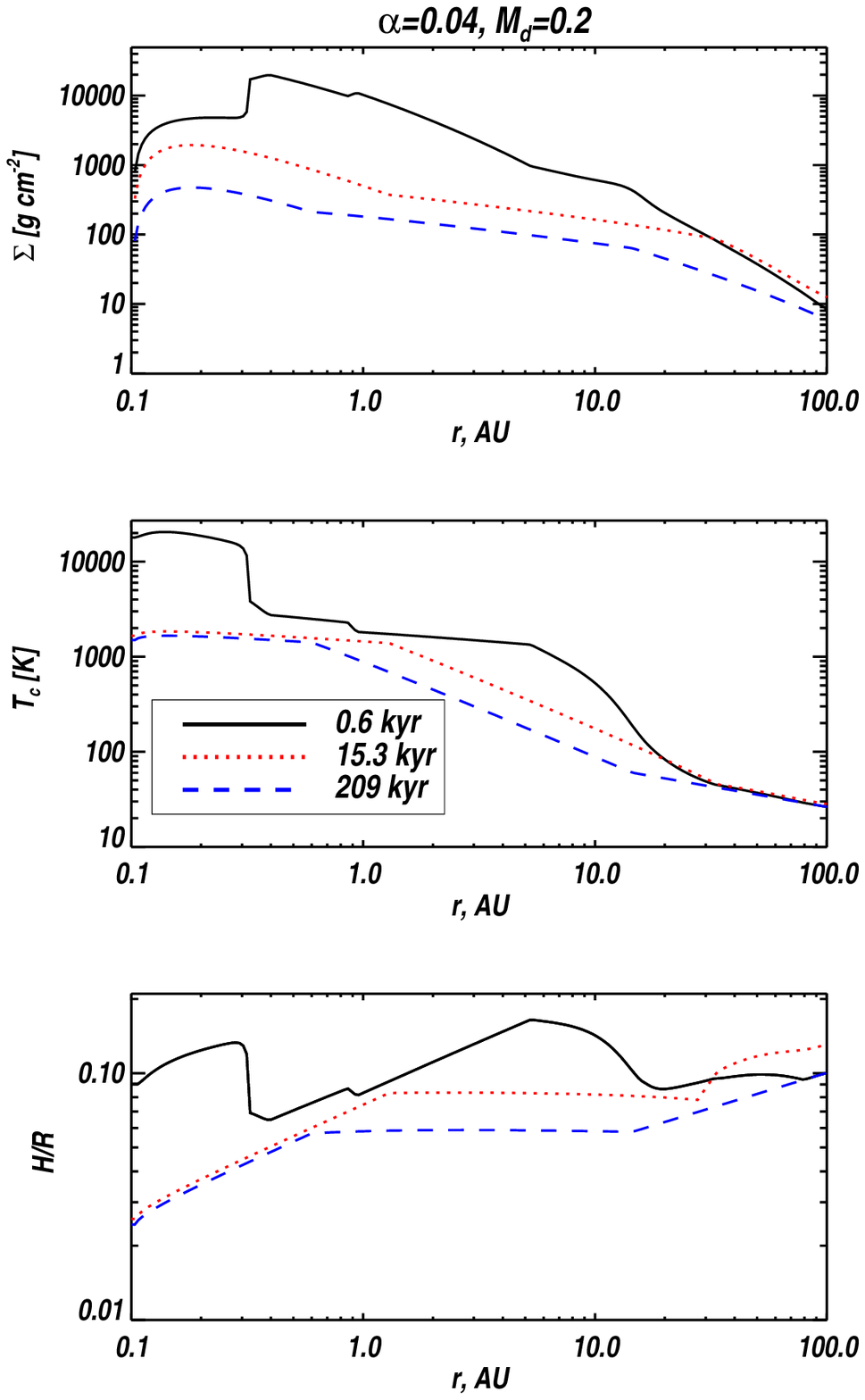,width=0.99\textwidth,angle=0}
\end{minipage}
  \caption{Disc evolution for two representative examples of a $1 \mj$ planet
    migrating in from the initial location of $75$~AU. The left hand side
    panels are for a low mass and viscosity disc, whereas the right hand ones
    are for a high mass and viscosity disc. The panels show disc surface
    density, $\Sigma(R)$, the central midplane temperature, $T_c$, and the
    geometric aspect ratio, $H/R$, from top to bottom, respectively. The disc
    profiles are shown at several times as shown in the legend.}
 \label{fig:two_discs}
\end{figure*}

Figure \ref{fig:two_discs} shows the disc evolution in two example
calculations for a planet of a fixed $M_p = 1 \mj$ mass migrating within a
protoplanetary disc, starting from $a(t=0) = 75$~AU. We are focused here on
the migration part of the model, so we assume that the planet has already
contracted to very high density and thus is not a subject to a tidal
disruption. The left hand panels of the figure shows the case of a relatively
low initial mass gas disc, $M_d = 0.04 \msun$, and a low viscosity parameter,
$\alpha_{SS} = 0.005$. The initial conditions for the disc are as described in
\S \ref{sec:IC}, except the disc inner boundary is set to $R_{\rm in} =0.1$~AU
in this section.

The right hand panels of figure \ref{fig:two_discs} contrast the low mass disc
case to a much more massive disc $M_d = 0.2\msun$ and a viscosity parameter
higher by a factor of 8, $\alpha_{SS} = 0.04$. Since the steady state
accretion rate in the disc scales as $\nu \Sigma$ \citep{Shakura73}, the right
hand panels' disc produces approximately 40 times larger accretion rate onto
the star than the left hand's disc. The sub-panels of the figure show the disc
surface density $\Sigma$, the central (midplane) disc temperature, $T_c$, and
the disc geometrical aspect ratio, $H/R$, at three different times as
indicated in the legend. The abrupt changes in $\Sigma$ and $T_c$ of the disc
in locations far away from the planet, e.g., at $R\approx 0.3$~AU, visible in
the black solid curves in the right hand panels are due to rapid dust and/or
gas opacity changes at these locations.

The first disc is much cooler, so that $H/R$ is smaller. Due to this, and
since the viscosity parameter is low, the disc already develops a deep gap
when the planet migrates to $a\sim 10$~AU. The gap remains deep but partially
opened until the planet migrates to $R\sim 0.6$~AU, at which point the inner
disc viscous time is shorter than the planet's migration time, so that the
inner disc drains onto the star and there is a complete gap between the star
and the planet. The planet continues to be "pushed" inward by the angular
momentum exchange with the outer disc, and eventually perishes by going
through the inner boundary of our computational domain for these tests,
$R_{\rm in} = 0.1$ AU.

In contrast, the planet is unable to open a gap of any denomination in the
second test because the disc is much hotter, so $H/R$ is higher, and because
the disc viscosity is high. This planet is always embedded in its parent disc
and migrates in type I regime only. The high disc viscosity in the second disc
implies that this disc evolves much faster. This can be surmised from the
$\Sigma(R)$ curves in figure \ref{fig:two_discs}: there is a notable decrease
in the disc surface density curves st late times even at large $R$ in the
middle right hand panel, whereas the corresponding curves on the left show
that the disc evolves only inside $R\simlt 10$~AU region for the low viscosity
test.

\begin{figure}
\centerline{\psfig{file=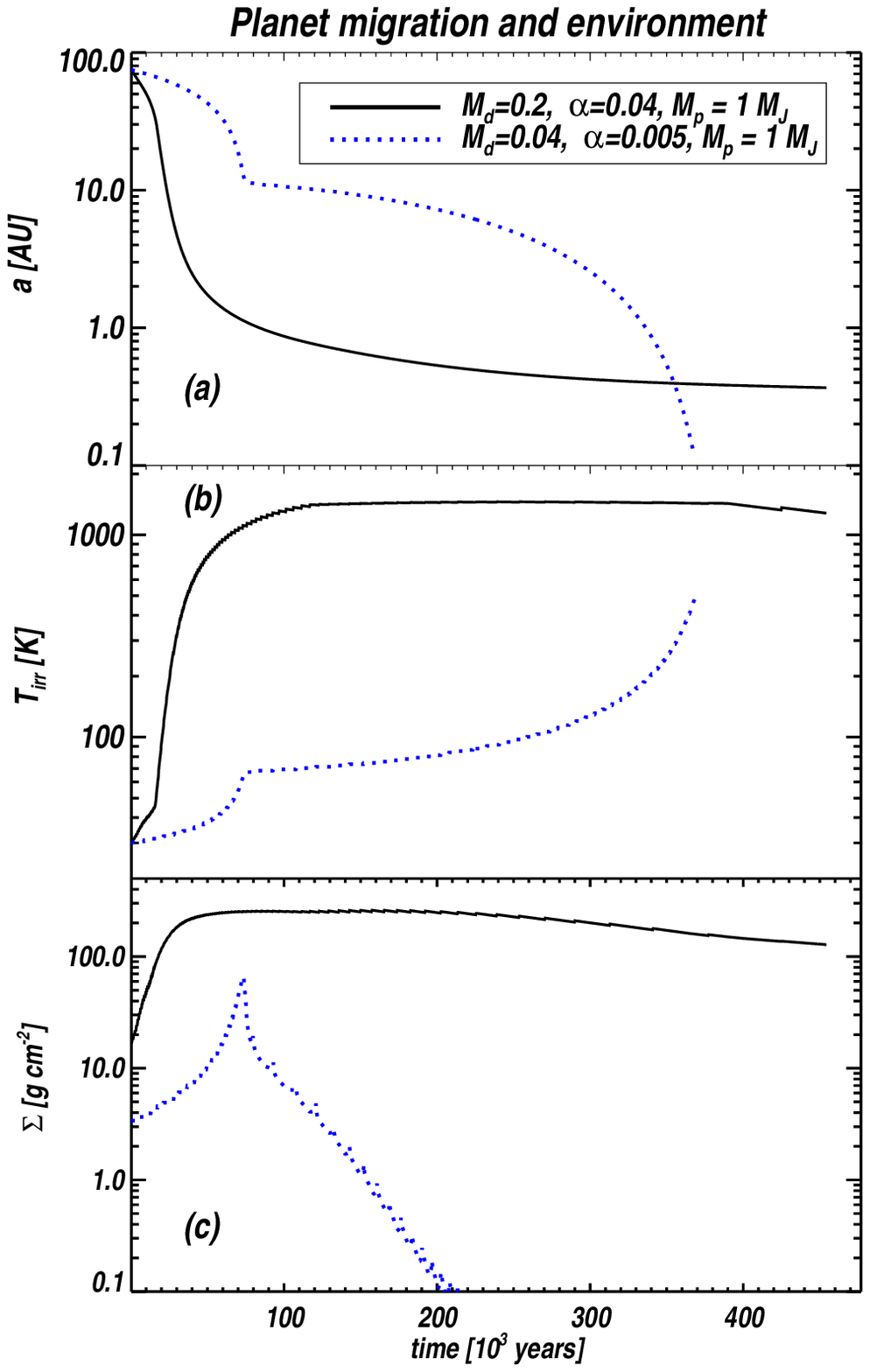,width=0.5\textwidth,angle=0}}
\caption{Planet separation versus time (panel a), irradiation temperature as
  seen on the surface of the planet (panel b), and the disc surface density at
  the location of the planet (c), for the two calculations presented in figure
  \ref{fig:two_discs}. Note that the hot massive disc planet (solid black
  curve) never opens a gap and always migrates in type I regime, whereas the
  other planet opens a deep gap when it reaches $a\approx 10$~AU at time
  $t\approx 70$~thousand years.}
\label{fig:two_discs2}
\end{figure}

Finally for this section, the time evolution of the disc environment around the
planet in the two calculations presented in figure \ref{fig:two_discs} is
plotted in figure \ref{fig:two_discs2}. Panels (a) through (c) show the
planet-star separation, $a$, the irradiation temperature incident on the
planet from the surrounding disc, $T_{\rm irr}$, and the disc surface density
at $R=a$, respectively. The initial evolution of these quantities, from
$a=75$~AU to $a\sim 10$~AU is similar for the two planets, modulo the fact
that the disc is five times less massive, and the planet migrates slower, for
the blue dotted curve than for the black solid one. At smaller distances to
the star, there is however a profound difference. While the hot disc planet
(solid curve) continues to migrate in the type I regime, initially quickly and
then slower and slower as it nears the inner disc, the other planet opens a
deep gap and starts to migrate in type II regime, as we already saw in figure
\ref{fig:two_discs}. Contrary to the hot disc case, the planet in the low mass
disc test accelerates as it enters the inner disc region. Therefore, somewhat
contrary to intuition, the less massive and cooler disc pushes the planet
through the inner boundary much faster (the hotter disc case planet arrives at
the inner boundary -- not shown in the figure -- about a Million years later).

Focusing now on the irradiation temperature evolution, two significant factors
should be noted. First of all, the planet in the hotter disc is bound to find
itself in hotter environment, generally. However, a secondary effect amplifies
this statement manyfold: since the planet in the hot disc does not open a gap,
it finds itself sampling the disc {\it midplane, e.g., central} temperature
$T_c$. The planet in the colder disc is in the gap, and that region is much
cooler: the appropriate irradiation temperature is comparable to the disc
effective temperature. There is a factor of at least a few difference in these
two temperatures {\it at the same radius $R$} because the disc optical depth
is usually considerable and hence the midplane temperature is higher than the
effective one ($T_{c} \approx T_{\rm eff} \tau^{1/4}$ in the Shakura and
Sunyaev 1973 disc theory, where $\tau = \kappa \Sigma/2 \gg 1$ is the disc
optical depth). In terms of irradiating flux incident on the planet, there is
thus a difference by many orders of magnitude. The "thermal bath" effect that
may puff up and eventually unbind a planet \citep[][and also section \S
  \ref{sec:bath} below]{CameronEtal82,VazanHelled12,DonWil14} is therefore much
stronger in the hot disc case than it is in the cold disc one.

These two tests illustrate why a detailed time-dependent calculation of disc
properties, and how these properties change due to the disc interaction with
the planet, is a necessity for an accurate statement on what happens with a
gas fragment migrating inward inside those discs.

\section{Planet evolution module}\label{sec:planet}

\subsection{Radiative contraction of the planet}\label{sec:rad_con}

\subsubsection{Pre-collapse evolution}\label{sec:pre-collapse}

\cite{Nayakshin10a,Nayakshin10b}, \cite{Nayakshin14b,Nayakshin14c}, employed a 1D
spherically symmetric Lagrangian radiative hydrodynamics (RHD) algorithm, with
grains modelled as a second fluid, to follow evolution of isolated
self-gravitating gas fragments.  Such a method is unfortunately too costly
numerically, and lends itself to studies of only a few test cases, while we
are interested in being able to investigate, eventually, thousands of models.

A more expedient "follow the adiabats" method
\citep[e.g.,][]{HenyeyEtal64,MarleauCumming14} is used in the present
paper. Namely, since the energy transfer inside the planet is strongly
dominated by convection at high opacities/metallicities \citep{HB11}, we
assume that the fragment is isentropic and is in hydrostatic balance. The
initial planet is assumed to have a constant grain to total mass ratio, equal
to the initial planet metallicity, $z$. The planet is divided into $N_b \sim
100$ concentric equal mass gas shells with mass $\Delta M = (1-z) M_p/N_b$,
where $M_p$ is the initial planet's mass. Given the central fragment's
temperature, $T_c$, at some time $t$, the structure of the isentropic planet
can be found by iterations on the central gas density, $\rho_c$, subject to
the condition that pressure vanishes at the outer boundary. This procedure
yields planetary radius, $R_p(t)$, the total energy of the planet, $E_{\rm
  tot}(t)$, and all the internal properties of the planet, e.g., the radial
coordinates for all of the radial mass zones within the planet, $R(M,t)$, as a
function of mass enclosed within the shell, $M$. In addition, the radiative
luminosity, $L(M)$, is calculated. $L(M)$ is small deep inside the planet,
increases outward and reaches a maximum value close to the planet's
surface. We take this maximum value to be the radiative luminosity of the
planet, $L_{\rm iso}$. This is done with the understanding that the very outer
regions of the planet will in reality be radiative rather than convective;
however the fact that the pressure scale height is usually very small compared
with the planet's radius means that the error in the value of planet's radius,
$R_p$, is also small.

The inner radius of the isentropic planet is given by $R_{\rm core}$, the
solid core radius, as described in \S \ref{sec:core}. We do not model the
internal structure of the core. The core releases $L_{\rm core}$, the core
accretion luminosity, into the rest of the planet.  The total energy of the
planet is evolved in time according to
\begin{equation}
{d E_{\rm tot}\over dt} = - L_{\rm rad} + L_{\rm core} - {G M_p
  \dot M_z\over R_p} \;,
\label{etot1}
\end{equation}
where $L_{\rm rad}$ is the radiative luminosity is the luminosity that the
planet would have in isolation, $L_{\rm iso}$, {\it minus} that incident on
the planet from the surroundings:
\begin{equation}
L_{\rm rad} = \max \left[0, L_{\rm iso} - L_{\rm irr}\right]\;.
\label{lrad1}
\end{equation}
Here $L_{\rm irr} = 4 \pi R_p^2 \sigma_B T_{\rm irr}^4$ is the irradiating
luminosity. $T_{\rm irr}$ is equal to the surrounding disc temperature if the
planet is embedded in the disc, or the local disc effective temperature if the
planet opens a gap, as explained in \S \ref{sec:2discs}. The last term on the
right hand side of equation \ref{etot1} is the change in the energy of the
planet if grains accrete on it at the rate $\dot M_z> 0$. Note that since
grains accrete on the planet slowly, that is at velocities not exceeding a few
m~s$^{-1}$ or else collisions would destroy the grains, the grains kinetic
energy input into the planet can be safely neglected \citep[see \S\S
  \ref{sec:gr_dyn} and \ref {sec:gr_growth}, and also][]{Nayakshin14d}.

\subsubsection{The thermal bath planet disruption}\label{sec:bath}

Equation \ref{lrad1} does not permit negative radiative luminosities for the
planet for the following reasons. The structure of strongly irradiated
planets, that is, when $T_{\rm irr} > T_{\rm eff}$, where $T_{\rm eff}$ is the
temperature of the planet in isolation, is not isentropic since the outer
layers become radiative \citep[e.g.,][]{BurrowsEtal08}. Experimenting with
strongly irradiated planets using the stellar evolution code MESA
\citep{PaxtonEtal11}, it was found that the outer layers of the planet become
approximately isothermal, $T\approx T_{\rm irr}$ \citep[see also the flat
  regions in $T$ vs pressure curves in figure 1 in][]{BurrowsEtal08}. The
radiative luminosity of the planet becomes much smaller than it is in
isolation. It does not become negative, however, except for a short time
required for the outer layer thermal balance re-adjustment. Furthermore, as
$T_{\rm irr}$ is increased, the depth of the outer radiative layer
increases. Importantly, when the layer contains a good fraction of the
planet's total mass and $T_{\rm irr}$ is comparable to the planet's virial
temperature, the outer layers start to expand without bounds until MESA stalls
by having to use tiny timesteps. Physically, the planet becomes unbound and is
destroyed very quickly. This is the "thermal bath" effect by which gaseous
protoplanets are known to be destroyed when the surrounding environment is too
hot \citep{CameronEtal82,VazanHelled12,DonWil14}.

The criterion that we use to capture the thermal bath destruction of the
planet is a qualitative one: we require the irradiation temperature to exceed
\begin{equation}
T_{\rm irr} \ge {1\over 2} T_{\rm vir}\;,
\label{bath}
\end{equation}
where the virial temperature of the planet is defined as $T_{\rm vir} = (1/3)
GM_p \mu/k_b r_p$, where $\mu = 2.4 m_p$ is the mean molecular weight.

\subsubsection{Post collapse ``hot start'' planets}

Gravitational potential of post-collapse planets is high enough to allow gas
accretion onto their surface without a need for radiation since H$_2$
dissociation energy can be such a sink. Hence gas accretion onto the planets
is much more likely in the post-collapse case than it is in the pre-collapse
case \citep[see also][]{NayakshinCha13}. Grains, including pebbles, would
accrete onto the post-collapse planet together with the gas, since such
accretion would inevitably generate very hot shock fronts where pebbles would
be easily vaporised and thus well mixed and coupled with the gas. However, gas
accretion onto planets is not included in this paper (see \S
\ref{sec:big_idea}) in an attempt to reduce the parameter space of the
models. Therefore our post-collapse planets do not accrete either gas or
grains and hence evolve at a constant mass unless disrupted.

To model the post-collapse contraction of the planets, publically available
code MESA \citep{PaxtonEtal11} is employed to create tables of radii, total
energy and luminosity for a given mass planet for {\it non-irradiated}
coreless giant planets at gas metallicity of $z=0.15$. This metallicity is
typical of the metallicity of the planets born in our discs due to metal
loading \citep{Nayakshin14c,Nayakshin14d}. The luminosity of the core is
neglected in the post-collapse phase because the core's heat content is always
much smaller than that of the post-collapse planets (unlike the pre-collapse
planet case).

Having $R_p(t)$ and $L_{\rm iso}(t)$ for isolated planets would be
completely sufficient for our purposes of deducing whether the planet
contracts more rapidly or gets tidally disrupted in the ``hot'' region
\citep[see the $a=0.05$~AU line in the top panel of figure \ref{fig:1}, and
  also][]{Nayakshin11a}, were it not for irradiation of the planet by the
surrounding disc or the central star. Due to that irradiation, the planet
contracts slower than an isolated planet would do. To take this into account,
we use an approach similar to that of the pre-collapse phase but employing the
MESA tables created and with core luminosity and pebble accretion turned off,
as explained above.  In this approach the energy equation becomes
\begin{equation}
{d E_{\rm tot}\over dt} = - L_{\rm rad}\;,
\label{etot2}
\end{equation}
where $L_{\rm rad}$ is calculated as in equation \ref{lrad1}. Now, having
found an updated value of the planet's total potential energy, interpolations
in the tables are employed to find the corresponding age of the planet and the
new values for $R_p(t)$ and $L_{\rm iso}(t)$. The contraction step can now be
repeated, and hence $R_p(t)$ for a migrated irradiated planet is found. 

The post-collapse planets can in principle be disrupted due to either tides
(\S \ref{sec:tides}) or the thermal bath effects (\ref{sec:bath}), but in the
present paper none of our planets go through such disruptions.


\subsection{Grain dynamics}\label{sec:gr_dyn}

Grain sedimentation is of course important only in the pre-collapse planets
since post collapse planets are too hot to permit existence of grains. In this
paper, grains are treated as a perturbation to gas dynamics. Having found
$R(M,t)$ and $R(M,t+\Delta t)$ as described in \S \ref{sec:pre-collapse}, we
find the gas velocity for every mass shell in the planet,
\begin{equation}
u(M,t) = {R(M,t+\Delta t)-R(M,t) \over \Delta t}\;.
\label{vdef1}
\end{equation}

If grains are sufficiently small, e.g., $a_g \simlt 0.01$~cm in practice,
where $a_g$ is a spherical grain's radius, then they follow the gas motion
closely due to a strong aerodynamic coupling. However, in general grain
dynamics can be different from that of the gas, therefore grain sedimentation
is modelled as in \S 3.4 of \cite{Nayakshin10a}. Grains are assumed to reach
their terminal sedimentation velocity (with respect to gas) quickly,
\begin{equation}
-u_{\rm sed} = (u_a - u) = - \frac{4 \pi G \rho_a a_g R}{3 c_s} \;\frac{\lambda+a_g}{\lambda}\;,
\label{vsed1}
\end{equation}
where $u_a$ is the grain velocity in the frame of the centre of the planet
(which is motionless in an isolated planet case but is moving with the planet
if the latter is embedded in the disc), $R$ is the radius of the given mass
shell inside the planet, $\rho_a$ is material density of the grain, $a_g$ is
the grain size, $c_s$ is the gas sound speed, and $\lambda$ is the mean free
path of $H_2$ molecules. Equation \ref{vsed1} joins smoothly the Epstein and
the Stokes drag regimes.

Grain sedimentation is opposed by turbulent grain mixing \citep[e.g.,][]{DD05}
and convection \citep[e.g.,][]{HB11}. These two processes can be combined to
give a diffusion coefficient, $D$, and be modelled as diffusion of grain
concentration, $\rho_g/\rho$ \citep{FromangPap06}. The equation describing
this process is
\begin{equation}
\frac{\partial (\rho_g/\rho)}{\partial t} = \frac{D}{R^2} \frac{\partial}{\partial
  R}\left[R^2 \frac{\partial (\rho_g/\rho)}{\partial R} \right]\;.
\label{diff_eqn}
\end{equation}
This defines mean grain diffusion velocity, $u_{\rm diff}$,
\begin{equation}
u_{\rm diff} = - D\frac{\partial}{\partial
  R}\left[{\rho_g \over \rho} \right]\;.
\label{vdiff1}
\end{equation}
Since the grain concentration, $\rho_g/\rho$, is highest in the planet's
centre due to grain sedimentation, the mean diffusion velocity is positive,
confirming that this process opposes grain settling. The diffusion velocity is
combined with the sedimentation velocity to give the total grain velocity in
the planet's centre of mass frame,
\begin{equation}
u_a = u - u_{\rm sed} + u_{\rm diff}\;.
\end{equation}
This defines the grain mass flux in or out of the Lagrangian gas mass
shells in the planet, which then defines the rate of change of the grain
density $\rho_g$ at every shell. Updating $\rho_g$ through these mass fluxes
is equivalent to solving the grain mass continuity equation,
\begin{equation}
{d\over dt} \left[R^2 \rho_g\right] = - \frac{d}{d R}\left[R^2 \rho_g
  u_a \right]\;.
\label{drhodt_full}
\end{equation}

The diffusion coefficient $D$ takes into account two effects: turbulence and
convection. For the former, a turbulence parameter $0 < \alpha_t\ll 1$ is
introduced,
\begin{equation}
D_{\rm turb} = \alpha_t R_p c_s\;,
\label{Dturb}
\end{equation}
where $R_p$ is planet's radius and $c_s$ is the local sound speed. Convective
contribution to $D$ is \citep{HS08} given by
\begin{equation}
D_{\rm conv} =  l_{\rm conv} v_{\rm conv}\;,
\label{Dconv1}
\end{equation}
where $l_{\rm conv}$ is the local gas scaleheight, and $v_{\rm conv}$ is the
speed with which convective eddies rise. On the basis of the mixing length
theory of convection \citep{KW90},
\begin{equation}
D_{\rm conv} =  \left[{F_{\rm conv} g \mu l^4\over 7 k_b \rho T}\right]^{1/3}\;,
\label{Dconv2}
\end{equation}
where $g = G M(R)/R^2$ is the local gravitational acceleration and $F_{\rm
  conv}$ is the local convective flux.  The two contributions are added
together linearly, so
\begin{equation}
D = D_{\rm turb} + D_{\rm conv}\;.
\label{Dtot}
\end{equation}
It should be emphasised that turbulence and convection are very important in
opposing grain sedimentation and therefore core's growth \citep{HS08,HB11}. If
$u_{\rm diff}$ is sufficiently large then grains cannot sediment and in this
case $\rho_g/\rho=$~const everywhere inside the planet to a good
approximation, as we shall see later on.

The procedure just described is in place for all of the three grain species,
independently of one another. This is necessary because the species have
different sedimentation velocities, grain size $a_g$, and are distributed
differently inside the planet.

\subsection{Grain growth}\label{sec:gr_growth}

Grains are expected to have a distribution of sizes, from interstellar values,
$0.005\; \mu\hbox{m} \simlt a_g \simlt 1 \mu$m \citep{MathisEtal77}, to
perhaps as large as a few cm by radius due to grain growth. A
collision-fragmentation model for dust grains is necessary to follow the grain
size distribution \citep[e.g.,][]{DD05,HB11}. The grain distribution function
is also a function of position inside the planet. Such a complicated treatment
is well beyond our numerical resources, given that grain growth is only one of
the important processes that is modelled here. In addition, physical
uncertainties in grain growth and fragmentation processes would require
introducing new poorly constrained parameters, greatly increasing the
parameter space for the population synthesis models.

Following \cite{Nayakshin10b,Nayakshin14b}, grain growth is captured in a
simpler framework. The grain distribution function over grain sizes is
approximated by just two components: the ``small'' and the ``large'' grains.
The small grains are always present due to fragmentation of larger grains in
collisions, as suggested by observations and modelling of grain growth in
proto-planetary discs \citep{DD05}. We assume that the small grains are
tightly bound to gas, do not sediment, and are, of course, the dominant source
of dust opacity.

The population of large grains is represented by a single size $a_g$ and is
allowed to move relatively to the gas as described above. The size $a_g$
should be thought as a typical radius of the large grain population in a given
gas mass shell. $a_g$ varies with time due to grain growth, fragmentation,
vaporisation, and motion of the grains from one radial cell to another.  These
processes are treated almost exactly as in \cite{Nayakshin14b}, and hence we only
give a brief summary of what is done and point out the differences.

Large grains grow due to Brownian motion of small grains, and by sticking
collisions as they sediment \citep[cf.][]{Boss98,Nayakshin10a}:
\begin{equation}
\left(\frac{da}{dt}\right)_{\rm grow} = \frac{\rho_g}{4\rho_a}  \left[u_{\rm
  Br}  +  u_{\rm sed} f_{\rm sb}(x)\right]\;,
\label{da_grow}
\end{equation}
where $u_{\rm Br}$ is Brownian velocity here set to 10 cm~s$^{-1}$ \citep[see
  \S 2.4 in][]{Nayakshin14b}, which is typically much smaller than the
sedimentation velocity. The ``stick-or-break function'' $f_{\rm sb}(x)$, where
$x= u_{\rm sed}/u_{\rm max}$, is given by
\begin{equation}
f_{\rm sb}(x) = 1- x
\label{fsb}
\end{equation}
This simplifies the treatment of grain collisions used in \S 2.6 of
\cite{Nayakshin14b}, but achieves the same physical goal: grain-grain
collisions are perfectly sticking for $u_{\rm sed} \ll u_{\rm max}$, but
become fragmenting when sedimentation velocity exceeds $u_{\rm max}$ (note
that $f_{\rm sb}(x>1)< 0$). The possible grain cross-fragmentation, e.g., Fe
grains fragmenting silicate grains \citep{Nayakshin14b}, is neglected in this
paper for simplicity.

The maximum velocity, $u_{\rm max}$, under which grain sticking is still
possible, depends on properties of the material, and is best inferred from
experiment. There is however a large uncertainty here. $u_{\rm max}$ was
measured to be a few m/s for both silicate
\citep{BlumMunch93,BlumWurm08,BeitzEtal11} and icy materials \citep{SA12}, but
\cite{DT13a} report first laboratory experiments on decimetre-sized dust
agglomerates that are bound by surface forces only (rather than by chemical
bonds important for materials such as gypsum), and find $u_{\rm max}$ as small
as $0.16$~m/s. Pure metallic Fe grains may have $u_{\rm max}$ as large as
$300$~m/s, although interstellar amorphous Fe grains are probably considerably
weaker \citep[see discussion in \S 2.6.1 in][]{Nayakshin14b}.  By performing
numerical experiments, it was found that $u_{\rm max}$ is an important
parameter of the model as it controls the maximum speed with which grains can
sediment, and hence the core can be assembled. It is thus left as a free
parameter of the models.

Grain vaporisation rate, $(da/dt)_{\rm vap}$, is calculated exactly
as in \S 2.5 of \cite{Nayakshin14b}. Grains vaporise rapidly when the
surrounding gas temperature exceeds the vaporisation temperature for given
species; the vaporisation temperature is a function of gas pressure.

Finally, due to diffusion and sedimentation grains may enter or leave a given
mass shell, which leads to a change in the grain size if $a_g$ is different in
different regions of the planet. Utilising the grain mass conservation
equation \ref{drhodt_full}, we write
\begin{equation}
\left({d a_g\over dt}\right)_{\rm adv} = - \frac{1}{r^2 \rho_g} \frac{d}{d
  r}\left[a_g r^2 \rho_g u_a \right]\;.
\label{ag_advect}
\end{equation}
This is an ``advective'' change in the grain size.  Combining all these
processes, we write the full equation for grain growth as
\begin{equation}
{da \over dt} = \left(\frac{da}{dt}\right)_{\rm grow} +
\left(\frac{da}{dt}\right)_{\rm vap} + \left({d a_g\over dt}\right)_{\rm adv}\;.
\label{dadt_full}
\end{equation}

Grain abundances are as in \S 2.8 of \cite{Nayakshin14b}, except silicates and
Fe are combined in one refractory species. The Solar abundance of metals is
defined as $z_\odot=0.015$ \citep{Lodders03}, which is divided amongst the
grain species such that the fractional abundances, $z_i$ of water, CHON and
silicates are 0.5, 0.25, 0.25 times metallicity $z$, respectively. For
material density of grains we use $\rho_a=3.5$~g~cm$^{-3}$, 1.5 and 1.0, for
silicates, CHON and water ice, respectively.

The processes of grain growth and sedimentation impose their own constraints
on the maximum timestep $\Delta t$, which may be more stringent than those
from \S \ref{sec:rad_con}. The timestep is required to be small enough that
none of the grain properties change by more than 10\% between any two
successful models of planetary structure.

\subsection{Core formation and growth}\label{sec:core}

The inner boundary of our computational domain is at the core's radius,
$R_1=R_{\rm core} = (3 M_{\rm core}/4\pi \rho_{\rm core})^{1/3}$, where
$\rho_{\rm core} = 3$~g/cm$^3$. This radius is many orders of magnitude
smaller than the outer radius of the planet, $R_p$. The structure of the
planet at such small radii is unresolved in our simulations. The boundary
conditions applied at $R_1=R_{\rm core}$ is $u=0$ and $L(R_1)=L_{\rm core}$.

As in \cite{Nayakshin10b,Nayakshin14b}, the core accretion rate depends on
velocity with which grains sediment, $u_a-u$, at the first radial zone above
the inner boundary,
\begin{equation}
\frac{d M_{\rm core}}{dt} = \psi(a_g)\cases{M_{d, 1} (u_{2}-u_{a2})/ R_{2}  \quad
  \hbox{if }\; {u_{2}-u_{a2}} > 0 \cr 0 \qquad \hbox{otherwise}\;.\cr}
\label{core_prescription}
\end{equation}
Here $M_{d, 1}$ is the mass of the dust in the first gas zone, $R_{2}$ is the
outer radius of the first gas zone, and $(u_{2}-u_{a2})$ is the velocity with
which grains arrive into the first zone from the second gas zone. 

The function $\psi(a_g)$ is introduced to quench growth of the core in the
cases where $a_g \ll 1$~cm, when grains should be coupled to gas
tightly. While the code reproduces the expectation that $d M_{\rm core}/dt$ is
very small for small $a_g$ (since $|u_{2}-u_{a2}|$ is very small), some
spurious core growth at small levels does occur. To turn this unphysical
growth off, the functional form of $\psi(a_g)$ is chosen to be
\begin{equation}
\psi(a_g) = \exp\left[ - \left({0.1\hbox{cm} \over a_g}\right)^2 \right]\;.
\end{equation}
This treatment is applied to all of the three grain species independently, to
determine the accretion rate of these onto the core. These accretion rates are
then added up to give the total core accretion rate. The core's mass is set to
a negligibly small value in the beginning of the simulations.

Energy release by the core may have important effects onto the rest of the
planet. The simplest approach would be to follow the CA prescription according
to which the core luminosity is 
\begin{equation}
L_{\rm core}^{(CA)} = {G M_{\rm core} \dot M_c\over R_{\rm core}} \;,
\label{lca}
\end{equation}
where $\dot M_c$ is the instantaneous accretion rate of the solids onto the
core. However, this approach is likely to strongly over-estimate $L_{\rm
  core}$ during its assembly {\em in TD framework} since this assumes that
radiation diffuses out of the core rapidly. This is probably wrong as opacity
of solid cores is very significant. The modelling of cooling of initially hot
rocky planets above Earth mass is still very uncertain due to insufficient
data on opacities, convection, conduction and other material properties at the
appropriate temperature and pressure ranges \citep[e.g.,][]{StamenovicEtal12}.
The atmospheres of such planets (the gas layers immediately adjacent to the
core) could hinder rapid cooling. For example, \cite{LupuEtal14} derive
cooling time scales for the atmospheres of Earth-like planets to be as long as
$10^5$ to $10^6$ years, whereas core assembly times in our models may be as
short as $\sim 10^4$ years.

Since a self-consistent treatment of the internal structure of the cores is
well beyond the scope of our work, a parameterisation of core accretion
luminosity is made following \cite{NayakshinEtal14a}, in which the core emits
its accretional energy on a finite time scale, $t_{\rm kh}$, where $t_{\rm
  kh}$ is the Kelvin-Helmholtz contraction time of the solid
core. Specifically,
\begin{equation}
L_{\rm core} = {E_{\rm core}\over t_{\rm kh}}\;,
\label{lcore1}
\end{equation}
where $E_{\rm core}$ is the residual potential energy of the core, which is
integrated in time according to 
\begin{equation}
{dE_{\rm core} \over dt} = {G M_{\rm core} \dot M_c\over R_{\rm core}} -
{E_{\rm core}\over t_{\rm kh}}\;.
\label{ecore2}
\end{equation}
$M_{\rm core}$ and $R_{\rm core}$ are the running (current) core's mass and
radius.  Evidently, if $t_{\rm kh} =\infty$, then $E_{\rm core}$ is exactly
equal to the potential energy of the core's assembly at mass $M_{\rm
  core}$. The core's luminosity is zero in this case. In more general case, if
$t_{\rm kh} \gg M_{\rm core}/\dot M_c$, then $L_{\rm core} \ll L_{\rm
  core}^{(CA)}$, but the total energy emitted by the core in the limit
$t\rightarrow \infty$ is the same ($GM_{\rm core}/2 R_{\rm core}$, where the
final value of the core's mass and radius are used). In the opposite limit,
when $t_{\rm kh} \ll M_{\rm core}/\dot M_c$, $L_{\rm core} = L_{\rm
  core}^{(CA)}$. In this paper, $t_{\rm kh} = 10^5$ years for all of the runs
presented. Preliminary results indicate that dependence of the results on
$t_{\rm kh}$ is moderately weak unless $t_{\rm kh}$ is longer than a few
million years, in which case most of the core's energy is saved "for later",
that is accumulated for release after the protoplanetary disc is removed.

\subsection{Tidal destruction of the planet}\label{sec:tides}

We assume that the gas envelope of the gaseous giant planet is {\it completely}
tidally disrupted when the planet fills a large fraction of its Roche lobe
radius, e.g., when
\begin{equation}
r_p > 0.7 r_H\;.
\label{disruption1}
\end{equation}
The factor in front of $r_H$ in this expression depends on the rotation state
of the planet. If the planet is in a synchronous rotation with the star, as is
usually the case for stellar binary systems \citep[e.g.,][]{Ritter88}, then
this factor is nearly unity. For hydrodynamical simulations of planets
embedded in discs, \cite{GalvagniMayer14} suggest the factor is $1/3$ whereas
\cite{ZhuEtal12b} find it closer to 0.5. These simulations however sample the
earliest phase in the evolution of the planets, some $\sim 10^3$ to perhaps
$10^4$ years into their existence. During this early phase the planets are
most rapidly rotating due to a significant angular momentum at formation
\citep[e.g.,][]{BoleyEtal10}, hence they may be quite aspherical and hence
easier to disrupt by tides. The factor of $0.7$ in equation \ref{disruption1}
is probably more relevant for the typically older planets that are studied
here. Our main results do not depend on this factor sensitively.

\subsection{Pebble accretion rate}\label{sec:dot_mz}

\cite{LambrechtsJ12} show that massive bodies accrete pebbles in the ``Hills
regime'', that is when all of the pebbles streaming past the planet within its
gravitational reach -- the Hills's radius $R_H = a(M_p/3 M_*)^{1/3}$ -- are
accreted. Therefore, 
\begin{equation}
\dot M_z = 2 f_{\rm p} \Sigma_g(a) v_H R_H
\label{dotmz}
\end{equation}
where $v_H = \Omega_a R_H$, $\Omega_a = (GM_*/a^3)^{1/2}$, and $\Sigma_g = f_g
\Sigma(R=a)$ is the grain surface density at radius $R=a$. The efficiency of
accretion of large grains by the planet is a function of their size
\citep[e.g.,][]{JohansenLacerda10,OrmelKlahr10}, and only a fraction $f_{\rm
  p}< 1$ is in the pebble regime.  $f_{\rm p}$ is a free parameter of the
model. 

Note that $\Sigma_g(a)$ must be found self-consistently by the disc-planet
interaction module (\S \ref{sec:disc}); it can vary by orders of magnitude at
the same distance from the star depending on the disc time evolution, and
especially on whether a gap around the planet's location is opened or not.

Having found the pebble mass accretion rate, pebbles are added to the existing
population of grains in the outermost few radial zones of the planet and are
then evolved in the manner described earlier.

\section{Combining the planet evolution and the disc migration modules}\label{sec:together}

The disc migration and planet evolution modules are combined together into a
time-dependent code. This is relatively straight forward. The two modules are
called by the main program that stores the values of all important variables
at a given time step and passes the information between the two modules.  In
particular, the disc migration module needs to be provided with the current
planetary mass, $M_p$, which changes relatively slowly due to pebble accretion
but can change abruptly due to a tidal disruption of the fragment. The planet
evolution module requires the following inputs from the disc migration routine: the
pebble accretion rate, $\dot M_z$, and the irradiation temperature, $T_{\rm
  irr}$. The main routine determines a time step, $\Delta t$, such that none
of the planet or the disc variables change by more than a few percent during that
step. If $\Delta t$ exceeds $\Delta t_{\rm max} = 10^3$ years (a free
parameter), it is set to $\Delta t_{\rm max}$. 

The migration and the planet evolution modules are then executed independently
from one another for the duration $\Delta t$. The exchange variables defined
above are held fixed during this step. After the execution they return the
information back to the main routine, which first makes a series of checks,
such as a comparison of the Hill's radius with the planet's radius to assess
the fragment's integrity against tidal disruption, and then repeats the
procedure by providing the modules with updated input parameters.

\subsection{Initial conditions}\label{sec:IC}

The disc is initialised with the following surface density profile:
\begin{equation}
\Sigma_0(R) = {A_m \over R} \left(1 - \sqrt{R_{\rm in}\over
  R}\;\right)\;\exp\left[-{R\over R_0}\right]\;,
\label{sigma_init}
\end{equation}
where $R_0$ is the disc length-scale, set to $R_0 = 120$ AU. The constant
$A_m$ is calculated so that the disc contains a given initial mass $M_{d} =
2\pi \int_{R_{\rm in}}^{R_{\rm out}} R dR \Sigma_0(R)$.  The initial surface
density profile given by equation \ref{sigma_init} is frequently used in
studies of protostellar disc evolution
\citep[e.g.,][]{MatuyamaEtal03,AlexanderEtal06}. The inner boundary of the
disc is a free parameter, set to $R_{\rm in} = 0.3$~AU for tests in \S
\ref{sec:opacity}. This is done for simplicity, since a detailed study of
planet evolution in the hot region $a\simlt 0.1$~AU is beyond the scope of
this paper. $R_{\rm out}$ is the outer radius of the disc
radial grid, and is set to a value significantly larger than $R_0$.

The planet is inserted at an initial location, $a=a_0\sim 100$~AU. The disc
structure around the planet initially does not take the planet's presence into
account, but this is not important in practice. The planet's mass is much
smaller than the surrounding disc mass at $a_0$ (unless an extremely low disc
mass $M_d \sim M_p$ is considered), and the planet is usually unable to open a
gap in the disc anyway until it migrates inward substantially, as found in 2D
simulations \citep{BaruteauEtal11}.

The planet's initial structure is a polytropic gas sphere of a given total
mass and central temperature, $T_c$, which is a free parameter of the
model. The grains are initially mixed with gas uniformly, i.e., at time $t=0$,
$\rho_g(R) = \rho(R) z_0/(1-z_0)$, where $\rho_g(R)$ is the total grain mass
density for all of the grain species, $\rho(R)$ is the gas mass density at
radius $R$ inside the gas clump, $z_0$ is the initial fragment's metallicity.
The volumetric density of a grain species $i$ is given by $\rho_{gi} = z_i
\rho(R)$. For simplicity, grains have a uniform initial size $a_0=10 \mu$m
everywhere in the cloud.

\subsection{Ending of the runs}\label{sec:end}

For a giant planet to survive the formation and disc migration phase, it must
stop migrating inward at some point. The may happen on the very inner edge of
the disc if the disc inner boundary is set by magnetospheric torques
sufficiently far from the stellar surface. Alternatively, the disc may
dissipate away due to photo-evaporation \cite[see][for a
  review]{AlexanderREtal14a} before the planet migrates all the way to the
inner boundary. Clearly, the balance between the rate of the disc
photo-evaporation and the planet's inward migration determines whether the
planet stalls or migrates to the disc inner edge. Our type I migration model
(\S \ref{sec:disc}) is too simplistic for the inner non self-gravitating part
of the disc, and therefore we do not attempt to end the simulations
``properly'', that is by photo-evaporating the disc \citep[e.g.,][]{AA09}. The
runs presented below are performed for just long enough for the planet to
migrate from the outer disc into the inner disc, and to thus answer the more
limited question of whether the giant planet-to-be survives this migration or
not. While a more self-consistent ending of the simulations is needed to
ascertain the eventual survival of the planets that collapsed before they
migrated into the inner disc, the present study remains useful at limiting the
spectrum of models that are able to overcome the tidal disruption barrier and
deliver giant planets into the inner disc.

%
%

\section{Low opacity models}\label{sec:opacity}

This section begins presentation of different tests of the model. We begin
with the case of no pebble accretion, $f_{\rm p}=0$. The input parameters and
main outcomes of selected simulations discussed in the sections to follow are
summarised in Table 1. The runs are labelled as "M1LowOpX" etc, where "X" is
an integer referring to a parameter varied within a given series of runs (see
below), and the preceeding letters refer to one of the three series of
runs. The number after "M" is the initial mass of the fragment in Jupiter
masses. All of the simulations are performed for the same disc parameters in
this paper, the disc viscosity parameter $\alpha_{SS} = 0.01$ and the initial
disc mass $M_{\rm d} = 0.12 \msun$. The conclusions of the paper are
qualitatively independent of the disc parameters. The main outcome of a
simulation -- fragment disruption or collapse -- is best inferred from the row
labelled $M_{\rm end}$, which shows the planet's mass at the end of the
simulation in Jupiter masses. For runs in which the planet is disrupted,
$M_{\rm end}$ is the mass of the solid core assembled inside the fragment
before it was disrupted, and it is of course much less massive than the
starting mass of the planet.

As we have seen in \S \ref{sec:big_idea}, the main challenge to forming giant
planets via TD is a long contraction time of giant planets, $\sim 1$~Million
years for $M_p = 1\mj$, which implies that planets migrate inward more rapidly
than they contract and hence they are disrupted \citep[see
  also][]{ZhuEtal12b,VazanHelled12,Nayakshin14c}. The migration time is found
to be $\sim 10^4 -10^5$ years, depending on disc parameters and the planet's
mass \citep[e.g.,][]{BoleyEtal10,ChaNayakshin11a,BaruteauEtal11}.

In order to get more giant planets to survive the fast inward migration, the
radiative contraction time of the planets is to be shortened somehow. One way
of achieving this is to invoke grain growth and therefore dust opacity
reduction \citep[e.g.,][]{HB11,Mordasini13}. To test this possibility, a
series of tests in which the opacity of \cite{ZhuEtal09} is multiplied by
arbitrary constant factors $f_{\rm op} = 10^{-(X-1)/3}$, where $X=1,2,..9$,
is ran. Jumping ahead, for $M_p=1\mj$, it is found that opacity reduction of
around 100 is needed for the planet to survive.

\subsection{Low mass fragments}\label{sec:low_mf}

\begin{figure}
\centerline{\psfig{file=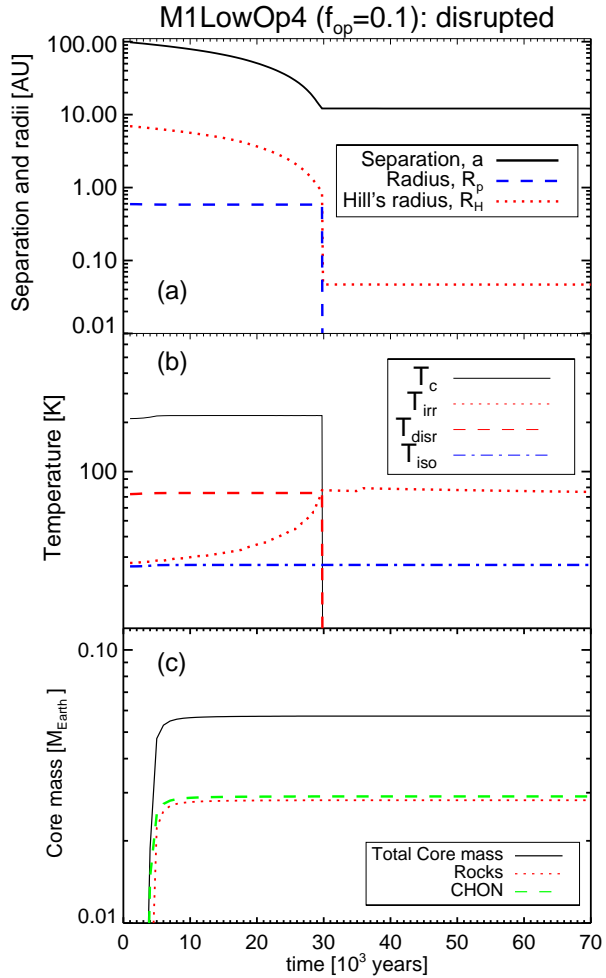,width=0.5\textwidth,angle=0}}
\caption{Simulation M1LowOp4 (see Table 1). ``Thermal bath disruption'' of a
  $M_p = 1\mj$ planet migrating from its birth location at $a=110$~AU occurs
  at $t\approx 30$~thousand years. Panel (a) shows disc planet separation,
  planet radius and the planet's Hill radius, as detailed in the legend. Panel
  (b) shows central temperature of the planet, $T_c$, irradiating, disruption
  and the effective temperature that the planet would have in isolation. Panel
  (c) shows the total core mass (solid curve) and how it breaks by the
  composition, as labelled in the legend.}
\label{fig:M1_op0.03}
\end{figure}

Figure \ref{fig:M1_op0.03} shows the simulation M1LowOp4 with $f_{\rm op}
=0.1$ ($X=4$). The upper panel (a) presents the evolution of the planet's
location, $a$, planet's radius, $R_p$, and the Hill's radius, $R_H$, with the
solid, dotted and dashed lines, respectively. The planet starts off at
$a=110$~AU and migrates inward on the time scale of $\sim 20,000$
years. Initially the planet is much more compact than the Hill's radius, but
$R_H$ shrinks as the planet's orbit shrinks whereas the planet's radius does
not. This eventually causes planet's disruption at $t\approx 30,000$~years.

The middle panel (b) of the figure shows several temperature characteristics
of the planet. In particular, the blue dash-dot line shows the effective
temperature of the planet in isolation, $T_{\rm iso}$, that is defined by
$L_{\rm iso} = 4 \pi R_p^2 \sigma_B T_{\rm iso}^4$, where $L_{\rm iso}$ is the
isolated planet luminosity (cf. equation \ref{lrad1}). The dotted red curve
shows the irradiation temperature. Since $T_{\rm irr} > T_{\rm iso}$ already
at birth of the planet, the planet cannot contract radiatively, since $L_{\rm
  rad}=0$ according to equation \ref{bath}. For this reason, the planet does
not contract; $T_{\rm c} =$~const and $R_p=$~const.

Since the planet continues to migrate inward rapidly, it is eventually
destroyed. As is seen in panel (b) of figure \ref{fig:M1_op0.03}, $T_{\rm
  irr}$ increases rapidly as the planet migrates in, and at $t=30,000$ years
the thermal bath planet destruction (\S \ref{sec:bath}) takes place since
$T_{\rm irr}$ exceeds $T_{\rm disr}$. In passing, it should be noted that if
the thermal bath destruction did not occur, this particular planet would have
been destroyed very soon nevertheless by tidal disruption since the tidal
disruption criterion (equation \ref{disruption1}) is almost satisfied at
$t=30,000$~years.

Panel (c) of Fig. \ref{fig:M1_op0.03} shows the mass of the solid core
assembled in the centre of the protoplanet. The remnant of the disruption is a
low mass solid core, $M_{\rm core} = 0.057 \mearth$. It is composed almost
equally of CHON and silicates. The presence of CHON materials in the core is
explained by the arrested evolution of the planet: since it is unable to cool
and contract, it stays relatively cold for longer, and this enables CHON
grains to sediment down into the core. Water ice cannot condense out even in
these conditions, which is consistent with earlier results of
\cite{ForganRice13b} who found that TD cores contain little water.

The remnant core is very low mass, so that it does not migrate appreciably
during the simulation, stalling at $a\approx 12$~AU. The simulations M1LowOpX
were all stopped at $0.3$ Million years, but the remnant's migration time is
$\sim 10^7$ years at the end of the run, so not much migration would occur in
the typical 3 Million years disc lifetime even if the simulations were
continued for longer.

\begin{figure}
\centerline{\psfig{file=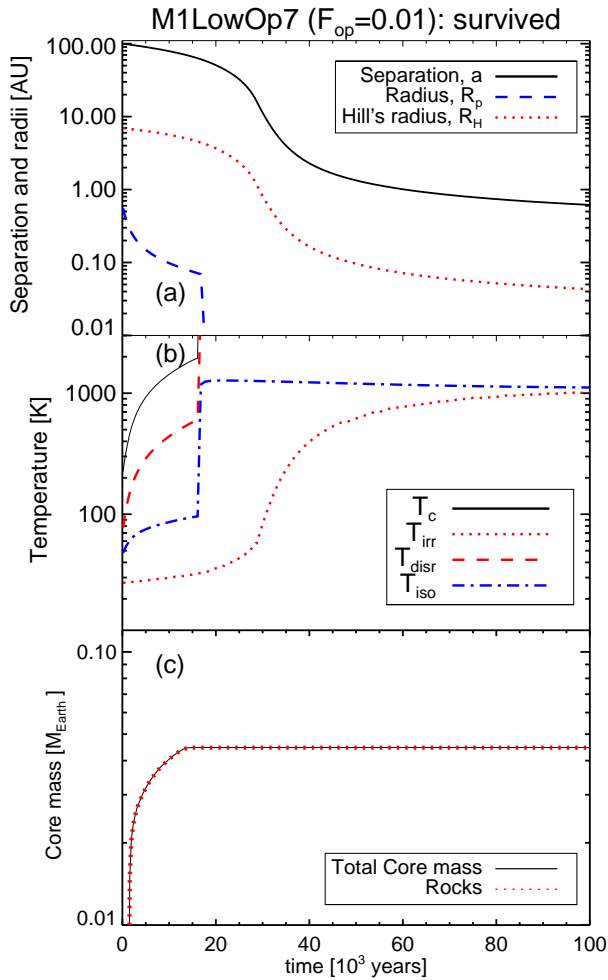,width=0.5\textwidth,angle=0}}
\caption{Same as figure \ref{fig:M1_op0.03} but for run M1LowOp7 (planet's
  opacity set to $f_{\rm op} = 0.01$). The planet radiative contraction is now
  faster than before and hence it is able to collapse before it is tidally
  disrupted.}
\label{fig:M1_op0.01}
\end{figure}

Figure \ref{fig:M1_op0.01} shows an identical run but with opacity further
reduced, $f_{\rm op} = 0.01$, labelled M1LowOp7 in Table 1. In this case, the
planet's intrinsic luminosity exceeds the irradiating one at early times
(cf. the dash-dot blue and the dotted red curves in panel b). The planet thus
contracts rapidly, and its central temperature increases rapidly (the solid
black curve in panel b). The planet undergoes the second collapse at $t\approx
20,000$ years, well before it is challenged thermally (due to external
irradiation) or tidally (due to tidal forces from the star). The core of the
planet is slightly less massive than in LowOp1, and is completely dominated by
silicates because the planet heats up quickly and the CHON grains have little
chance to sediment into the core.

Although we do not present the run M1LowOp6, for which $f_{\rm op} \approx
0.022$, we note that it ended with the planet disrupted. The conclusion from
this series of tests is that opacity reduction by a factor of $\sim 100$,
compared to the interstellar grain opacity at Solar metallicity, is needed to
counter the short inward migration times and the thermal bath effect, at least
for the parameters of the disc and the planet mass chosen in this section.

Figure \ref{fig:Mcores} shows the mass of the cores formed in the series of
runs M1LowOpX as a function of the opacity reduction factor $f_{\rm op}$, and
section \ref{sec:discussion} considers broader implications of these runs for
TD theory of planet formation.

\subsection{Higher mass fragments}\label{sec:high_mf}

\begin{figure}
\centerline{\psfig{file=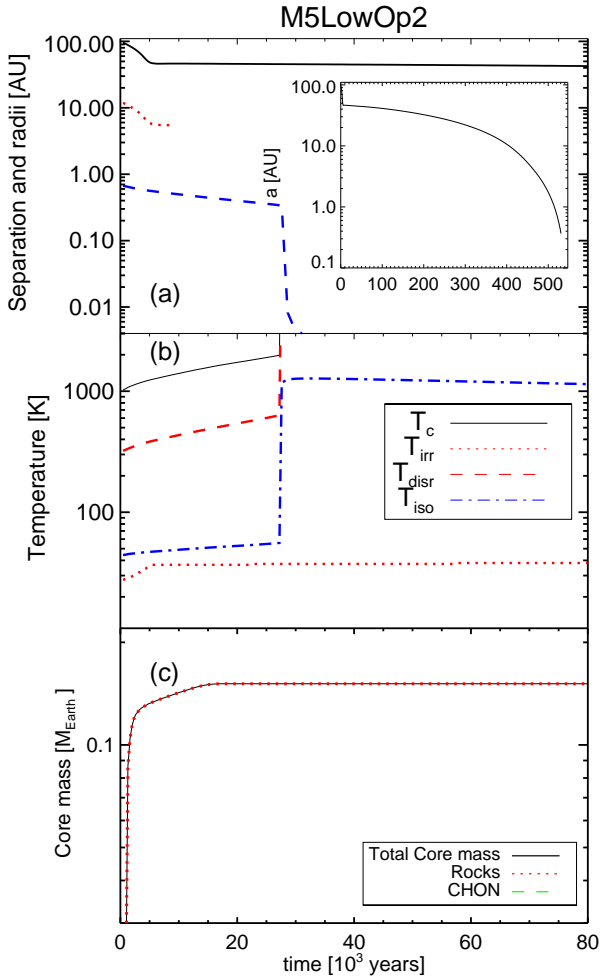,width=0.5\textwidth,angle=0}}
\caption{Same as figure \ref{fig:M1_op0.03} but for simulation M5LowOp2, e.g.,
  a more massive gas fragment, $M_p = 5 \mj$, and higher dust opacities
  ($f_{\rm op} = 0.46$).}
\label{fig:M5_op0.1}
\end{figure}

Finally for this section, figure \ref{fig:M5_op0.1} shows simulation M5LowOp2,
where the fragment is more massive, $M_p =5 \mj$. The practical interest in
considering more massive fragments is that such fragments cool much more
rapidly than $\sim$ Jupiter mass ones \citep[see figure 1
  in][]{Nayakshin14c}. Therefore, one may reason that dust opacity may not
need to be reduced by orders of magnitude for such fragments.

Indeed, in simulation M5LowOp2 (see Table 1), $f_{\rm op} = 0.46$, that is,
the dust opacity is reduced by only a factor of 2, yet the planet is able to
contract and collapse before being tidally disrupted. The collapse occurs at
$t\approx 27,000$ years, when the planet is at separation $a\sim 46$~AU. One
reason for the fast contraction of the planet is that the planet's effective
isolated temperature, $T_{\rm iso}$, is always above the irradiation
temperature (cf. panel b), and the second is that the planet's migration is
dramatically slowed down when the planet opens a gap in the disc at $a\sim
45$~AU.  However, the planet eventually migrates all the way to the inner
boundary condition at $R=0.3$~AU after about 0.5 Million years (see the inset
in panel a of the figure). It would become either a hot jupiter, if its
migration stalls in the regions closer to the star that are not simulated
here, or be swallowed by the star completely.

Note also that despite having 5 times more mass in metals than the $M_p=1\mj$
runs, the fragment only manages to assemble a tiny silicate core, $M_{\rm c} =
0.15\mearth$ (cf. panel c of the figure and the Table). This is because the
fragment spends too little time in the molecular H regime (the pre-collapse
stage), so that even the silicate grains have too short a time window in which
to grow and sediment. This finding is consistent with results of
\cite{HS08,HelledEtal08} who found that the higher the mass of the fragment,
the less efficient is core formation.

To save space, the simulation M5LowOp1 is not shown in figures but its results
are listed in Table 1. In this higher opacity, $f_{\rm op}=1$ case, the planet
does not manage to contract into the giant planet and is disrupted at
$a=6.7$~AU. The core formed in this simulation is only slightly more massive,
$M_{\rm core} = 0.18\mearth$.

From the M5LowOpX series of runs we learn that (i) the planet must be more
massive than about $5-6 \mj$ to contract faster than migrate in at Solar
metallicity, at least for the disc parameters chosen in this series of runs,
and that (ii) massive planets do not make much more massive cores than less
massive planets do. This last finding is not new at all \citep{HelledEtal08}.

\section{Pebble Accretion models}\label{sec:pebbles}

\subsection{Giant planet survival}\label{sec:peb_sur}

In this section, models with pebble accretion onto the gas fragments and with
interstellar dust opacity, scaled to the instantaneous metallicity of the
fragment, are considered. The initial conditions for the disc and the planet
are exactly the same as for the runs M1LowOpX, but here the opacity reduction
factor is $f_{\rm op}=1$. Additionally, since the metallicity of the fragments
now varies with time as pebbles accrete onto the fragments, the opacity
coefficient is chosen to be proportional to the fragment's metallicity:
$\kappa = \kappa_0(\rho,T) (z/z_\odot)$, where $\kappa_0$ is the interpolated
table dust and gas opacity at Solar metallicity from \cite{ZhuEtal09}. Pebble
accretion rate is calculated as described in \S \ref{sec:dot_mz}.

Table 1 lists input parameters and key results from several runs of this
series, which are labelled M1PebX, where X is an integer between 1 and 9 which
sets the pebble mass fraction in the disc, $f_{\rm p} = 0.5 \times
10^{-(X-1)/4}$. Presumably, higher metallicity discs would have higher $f_{\rm
  op}$ as grain growth is faster at higher metallicities, so high (low) values
of $f_{\rm op}$ may be taken as a proxy for high (low) values of disc
metallicity.

\begin{figure}
\centerline{\psfig{file=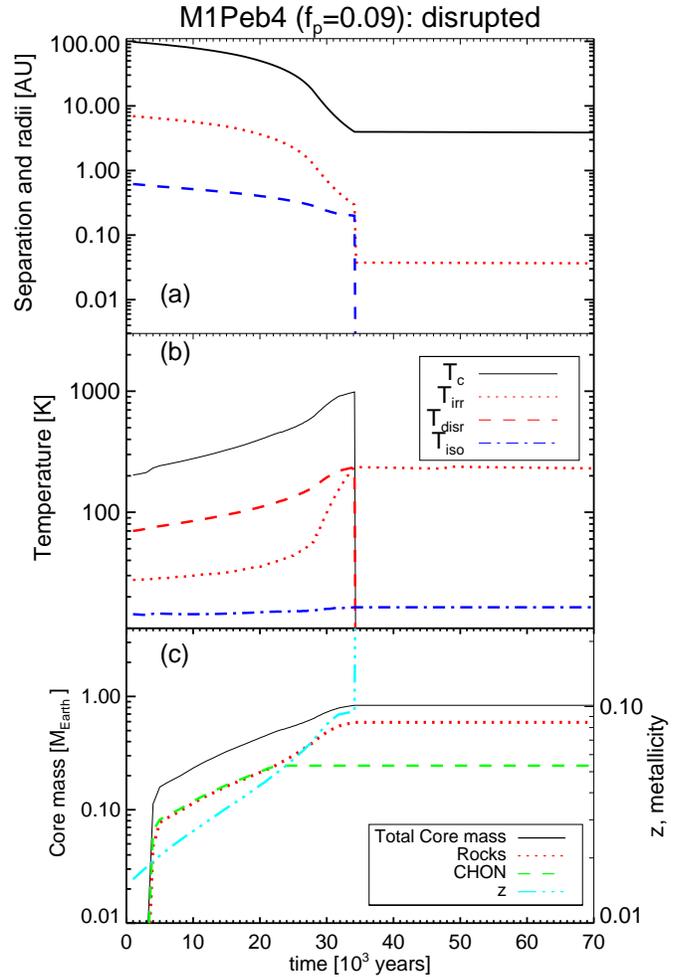,width=0.5\textwidth,angle=0}}
\caption{Same as figure \ref{fig:M1_op0.01} but assuming interstellar dust
  opacity ($f_{\rm op} = 1$) and allowing for pebble accretion with pebble
  fraction in the disc $f_{\rm p} = 0.09$ (run M1Peb4 in Table 1). This
  experiment ends in giant planet destruction and a remnant of $0.84 \mearth$
  masses.}
\label{fig:M1_terr1}
\end{figure}

Figure \ref{fig:M1_terr1} shows the results of the run M1Peb4, $f_{\rm p} =
0.089$ in the format similar to figure \ref{fig:M1_op0.01} to
\ref{fig:M5_op0.1}, except that the bottom panel (c) also shows the evolution
of the planet's metallicity, $z$, defined as the ratio of the mass of the
metals to the total mass of the planet. $z$ is shown with the cyan
dash-triple-dot curve, and the relevant scale is given on the right hand side
edge of panel (c).

Due to pebble accretion, the planet's metallicity increases with time sharply
until time $t\approx 30,000$~years. At that point accretion of pebbles slows
down since the planet starts opening a gap in the disc around its location,
so that $\Sigma_g(a)$ plummets. As a result, the planet contraction slows down
(note that the dashed blue curve in panel (a) flattens at that point). Due to
a continuing inward migration, the Hills radius of the planet (red dotted
curve in panel a) continues to shrink, and the planet is tidally disrupted at
time $t= 34,000$~years, when the tidal disruption criterion (equation
\ref{disruption1}) is met.

The core's
mass at that point is $M_{\rm core} = 0.84\mearth$ and the planet's location
is $a=3.97$~AU. Since the core's possible atmosphere is neglected in this
paper, the mass of the planet after the disruption is set to that of the core,
and the metallicity is set to $z=1$ by definition. Also, the temperature of
the planet is arbitrarily set to 0 to make the cases of envelope disruption
visibly distinct from the cases when the envelope collapses and heats up
(e.g., fig. \ref{fig:M5_op0.1}). In reality the core's temperature may be
quite high but that is not modelled here (\S \ref{sec:core}).  The core's
composition in run M1Peb4 is dominated by silicates (see panel c of
fig. \ref{fig:M1_terr1}), yet over a third of the core's mass comes from
organics (CHON).

Next figure, \ref{fig:M1_hjup1}, shows simulation M1Peb3, which has a slightly
higher fraction of pebbles in the disc, $f_{\rm p} = 0.16$. This turns out to
be sufficiently high to enforce collapse of the fragment before it is tidally
or thermally disrupted. Metallicity of this planet at collapse is
$z=0.152$. This higher metallicity translates into larger grain sizes in the
planet (since $(da/dt)_{\rm grow}$ term in equation \ref{dadt_full} is larger)
and a higher grain growth rate of the core. The mass of the core at collapse
is $M_{\rm core} = 3.65\mearth$. It is interesting to note that the CHON mass
of the core is actually slightly lower than in the run M1Peb4, and hence the
core is over 90\% silicates. This result is driven by a faster contraction
rate of the planet in M1Peb3. The time window for CHON grains settling into the core is now
shorter, so despite more CHON mass inside the planet, a smaller mass of that
reservoir arrives at and is locked into the core.  The planet in simulation
M1Peb3 continues to migrate in a regime intermediate between type I and II
regimes and eventually arrives at the disc's inner edge. The planet may
survive as a hot jupiter if the disc is removed while the planet is still
migrating in or if the inner boundary of the disc is cut off by magnetospheric
interactions sufficiently far from the star's surface.

Finally, Table 1 lists the results of the most pebble-rich run of the series,
M1Peb1, $f_{\rm p}=0.5$. Interestingly, the higher abundance of pebbles did
not increase the core's mass or the planet's metallicity at collapse. The
latter is due to the fact that the planet collapses when it accretes enough
pebbles \citep{Nayakshin14c}, which is independent of the rate at which the
metals are added to the planet. The core's mass appears to be lower due to a
shorter time span available for grain sedimentation.

In summary, runs M1PebX with X=1,2,3 produced a giant planet that collapses
before it was tidally disrupted, whereas runs with X$\ge 4$ resulted in tidal
or thermal disruptions of the giant planets-to-be. Since a higher pebble
accretion rate may be expected at higher metallicity environments, one expects
a positive correlation of the giant planet detection frequency with
metallicity of the host, as shown by \cite{Nayakshin14d}.

\begin{figure}
\centerline{\psfig{file=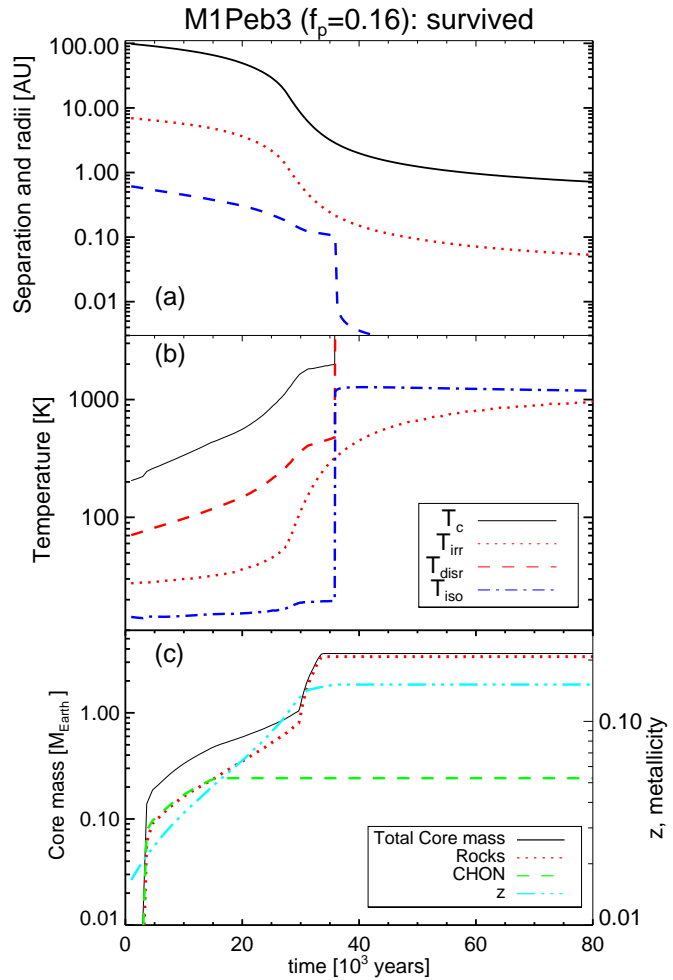,width=0.5\textwidth,angle=0}}
\caption{Same as figure \ref{fig:M1_terr1} but a higher pebble fraction,
  $f_{\rm p} = 0.16$ (M1Peb3 in Table 1). Unlike M1Peb4, the giant planet
  manages to contract and collapse in this case. The end result is a metal
  rich, $z=0.15$, hot jupiter with a core mass of $3.65\mearth$.}
\label{fig:M1_hjup1}
\end{figure}

\subsection{Planet's internal structure}\label{sec:internal}

\begin{figure}
\centerline{\psfig{file=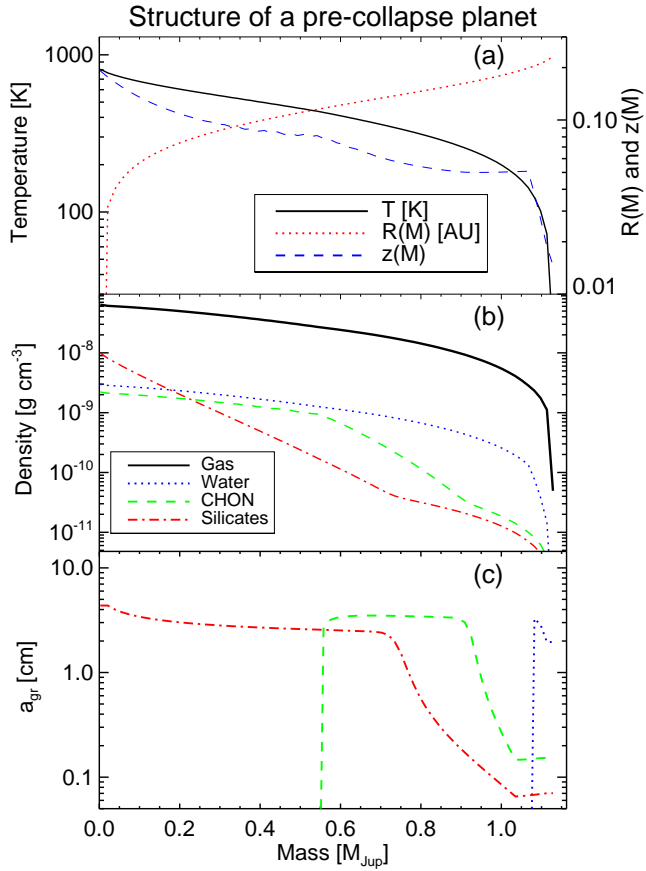,width=0.5\textwidth,angle=0}}
\caption{Internal structure of the planet at time $t=24450$~years in simulation M1Peb3 as
  a function of total (gas plus metals, including the core) enclosed
  mass. Panel (a) shows the temperature, Lagrangian radius (in units of AU), and local
  metallicity, $z(M)$. Panel (b) shows gas (solid) and the three grain metal species
  density profiles, while panel (c) shows the species' grain size, $a_{\rm
    gr}$. }
\label{fig:structure}
\end{figure}

Figure \ref{fig:structure} shows the planet's internal variables as a function
of the total (gas plus metals) mass enclosed in concentric shells for a planet
from run M1Peb3 at time $t= 24450$ years, before it is tidally disrupted. The
planet's structure in the pre-collapse (or pre-disruption) stages are similar
to one another, so Fig. \ref{fig:structure} is representative of the pre-collapse TD
planets structure in general.

The temperature in the planet, as expected, is maximum in the
centre and falls off towards the planet's outer edge; see the solid curve in
panel (a). In the same panel, the dotted red curve shows radius in units of AU as a
function of the enclosed mass. The planet's outer radius is $R_p\approx
0.22$~AU. The blue dashed curve shows the local metallicity, $z(M)$, defined
as the ratio of the metal's mass in the given mass shell to the total mass of
that shell. 

Note that the highest metallicity gas is at the centre of the planet, as may
be expected if metals (grains) are able to sediment to the planet's centre. In
addition, there are two somewhat sharp features in the $z(M)$ function, one
at $M\approx 0.55 \mj$, and the other near the outer edge of the planet. These
two features mark two important transitions within the planet. The nature of these transitions is best inferred from the bottom panel of Fig. \ref{fig:structure}, (c), which
shows the grain size for the three grain species that are considered in this
paper: water, CHON and silicates. The vaporisation temperatures for these
species are very different (see figure 1). 

The silicate grains are the most refractory of the three, so they are able to
sediment down all the way to the centre of the cloud in Fig. \ref{fig:structure}. 
Despite this, note that the grain size
for silicates is limited to a few cm. This is a direct consequence of the
grain breaking velocity set to $v_{\rm b} =10^3$~cm/s for this
  simulations. Larger grains would sediment faster than this and would
  fragment due to collisions. In the outer regions of the planet, the grains
  are not limited by fragmenting collisions. Instead, the limitation here
  comes from the balance between the rate at which the grains grow and the rate
  at which they sediment down: larger grains are constantly removed from the
  outer regions of the planet by sedimentation, so there exist a quasi equilibrium
  between small growth injection at the outer edge of the cloud, grain growth and sedimentation of the grains.

Grain diffusion due to convection in particular is a very important part of
this quasi-equilibrium, since convection opposes grain sedimentation
\citep[e.g.,][]{HB11}. Effects of convective grain mixing are best revealed by
looking at the grain volume density distributions (the middle panel, b). It is
notable that there are transitions in $d(\rho_{i}/\rho)/dm$, where $\rho_i$ is
the grain volume density of species $i$, at locations where grain size $a_i$
changes abruptly. For example, CHON grains are ``small'' in the outer region
of the cloud, then ``large'' between $M\approx 0.9\mj$ and $M\approx 0.55\mj$,
and very small again for $M\simlt 0.55 \mj$. Just as with silicate grains,
this is because grain growth is sedimentation-limited in the outer region,
then fragmentation-limited in the middle, and finally grain vaporisation
dissolves the grains and puts them into the gas phase. Convection
controls CHON grains' volume distribution in the inner part of the cloud, where
$\rho_{\rm chon}/\rho =$ const to a good degree. In the fragmentation-limited
range, $\rho_{\rm chon}/\rho$ is strongly decreasing outward, until the grains are
small and convection again takes over. Water ice grains are able to settle
down only through the outermost $\sim 0.1\mj$ of the planet, and therefore
there is a very large gradient in the water ice concentration there. In the rest
of the planet water is very well mixed with the H/He phase.

These grain growth, fragmentation and convective mixing equilibria drive the
chemical abundances of the species with the planet. In the outer $\sim 0.3
\mj$ of the planet, water is more abundant than CHON and silicates by a factor
of $\sim 30$ rather than the expected (Solar) abundance ratio (two). On the
other hand, silicates are over-abundant over water and CHON compared to their
initial relative abundances by a factor of $\sim 10$ in the centre of the
planet. TD fragments are hence natural thermo-mechanical element
differentiation ``factories'' while they are in the molecular H (pre-collapse)
phase \citep[see][for more on this point]{Nayakshin14b}.

\section{Discussion: low opacity or pebble accretion to save giants in TD?}\label{sec:discussion}

\begin{figure}
\centerline{\psfig{file=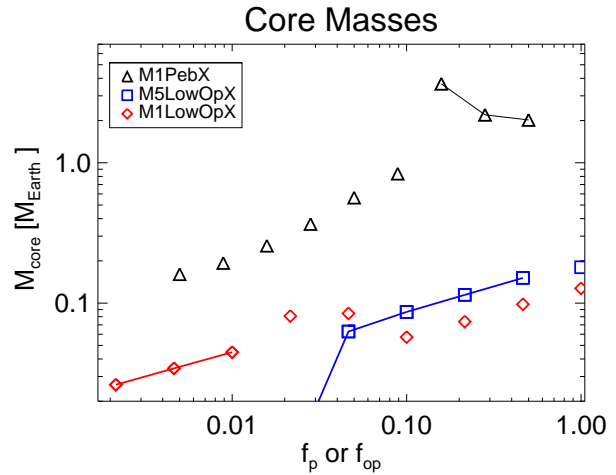,width=0.5\textwidth,angle=0}}
\caption{Core masses as a function of opacity reduction factor $f_{\rm op}$ or
  pebble fraction $f_{\rm p}$ for the three sets of models summarised in Table
  1. The runs for which gas fragments collapsed faster than they migrated in
  are connected with lines. For symbols not connected with lines, the
  fragments were tidally disrupted and their cores released back in the disc.}
\label{fig:Mcores}
\end{figure}

Figure \ref{fig:Mcores} presents a bird's view of the results of the three
series of runs presented in Table 1 by showing only the dependence of the core
mass, $M_{\rm core}$, on the pebble mass fraction, $f_{\rm p}$, for the pebble
accretion runs M1PebX (triangles), and on the opacity reduction factor,
$f_{\rm op}$, for the "low opacity" runs M1LowOpX or M5LowOpX (diamonds and
squares, respectively). In addition, the runs that resulted in the collapse of
the giant planet rather than its disruption are connected with one another by
solid segments of the same colour as the symbols.  For example, the three
black triangles in the upper right of Fig. \ref{fig:Mcores} show the runs
M1Peb1 to M1Peb3, for which the pebble mass fraction is the highest. The rest
of the black triangles refer to runs M1Peb4 to M1Peb9, for which the fragments
were disrupted before they could collapse.

It is of course
expected that $f_{\rm p}$ and $f_{\rm op}$ are proportional to disc metallicity,
since both are expected to increase as $z$ increases, with other parameters being equal.

We see that opacity reduction by a factor of $\sim 100$ compared to the
interstellar opacity at Solar metallicity is needed for planets of 1 Jupier
mass ( red diamonds) to collapse sooner than they are disrupted. 
It could be argued that planet migration inside lower mass discs than $M_d = 0.12 \msun$
used universally for all the tests in Table 1 could allow for a smaller
reduction in opacity (that is, higher $f_{\rm op}$), but much lower mass discs are
unlikely to become self-gravitating and give birth to a GI fragment in the
first place.

Furthermore, observed giant planets in the Solar System and beyond
\citep[e.g.,][]{MillerFortney11} are known to be over-abundant in metals by a
factor of $\sim 10$ or more than $z_\odot$. Thus the needed opacity reduction
is a factor of $\sim 10^3$ for the observed metal-rich giant planets of about
one Jupiter mass. While grain growth can reduce opacity
\citep{HB11,Mordasini13}, grain fragmentation should be included in such
models as well \citep{DD05}. We have seen that collisions with grains as large
as a few cm in size are frequent enough inside the pre-collapse fragments (see
\S \ref{sec:internal}) to limit grain size by fragmentation. This was found
for the breaking velocity set to $v_{\rm br} =10$ m~s$^{-1}$, a relatively
high value, but perhaps reasonable for the refractory grains strengthened by
sintering in the high temperature environment inside the fragment
\citep[e.g.,][]{Nayakshin14b}. Amorphous grain aggregates are found to
fragment at velocities as little as $\sim 0.16$ m~s$^{-1}$. Despite these
uncertainties in the input grain physics, it is hard to imagine that
fragmentation of smaller grains would not be efficient inside the pre-collapse
planets. It appears unreasonable to us to require the dust opacity to be
reduced by such a huge factor as $\sim 10^3$.

Even if nature does manage to reduce dust opacity to such tiny values somehow, the
trends of the low opacity TD models for giant planet formation directly
contradict the observations. It is obvious from fig. \ref{fig:Mcores} that low
metallicity environments would be more hospitable to planet formation via
GI/TD models {\it if} radiative cooling was the bottle neck for giant planet
contraction, in agreement with the results of \cite{HB11}. This trend is exactly
opposite to what is observed \citep{FischerValenti05}.

Furthermore, as seen from fig. \ref{fig:Mcores}, the core masses assembled by
such rapidly collapsing gas fragments would be rather small, $M_{\rm core}\sim
0.1\mearth$. When the "low opacity" run fragments (diamonds in
Fig. \ref{fig:Mcores}) are tidally or thermally disrupted, the mass of their
cores is far too small to explain rocky terrestrial planets, save for the more
massive super-Earth planets.

Considering more massive gas fragments relaxes the requirement for the low
opacity, as is seen from the sequence of squares in fig. \ref{fig:Mcores}. For
$M_p =5 \mj$, the required reduction in opacity is only a factor of $\sim 2$,
which is quite reasonable. However, the mass of the cores assembled inside such
high mass planets is still very low. This is because more massive planets are
hotter to begin with and contract more rapidly, leaving too little time for
grains to settle down \citep[as was found previously
  by][]{HelledEtal08,Nayakshin10b}. Disruptions of high mass fragments would
therefore fail to explain the abundant massive rocky planets
\citep[e.g.,][]{PetiguraEtal13}. The high mass gas fragments would also not be
able to explain formation of lower mass giant planets, $M_p\simlt 1-2 \mj$,
which are much more abundant in the data than planets of mass $M_p \simgt 5
\mj$. There is also no clear reason why the high mass radiatively cooling
fragments would follow a positive metallicity correlation.

Finally, whatever the mass of the gas fragment, the low opacity models do not
provide an explanation for why giant planets are found to be much more
abundant in metals than their host stars.

Therefore, it is our opinion that in terms of 
development of TD planet formation theory that could account for all types of
observed planets, low dust opacity models are not only physically unlikely,
they are a dead end.

In contrast, pebble accretion models (i) yield a positive planet-metallicity
corelation \cite{Nayakshin14d}; (ii) are very metal rich, e.g., $z\simgt
0.1-0.2$ (cf. entry $z_{\rm end}$ in Table 1) ; (iii) can assemble
sufficiently massive solid cores (cf. fig. \ref{fig:Mcores}) to potentially
explain rocky planets as remnants of disrupted gas fragments
\citep{BoleyEtal10}. While there is a number of issues not yet addressed by
TD, it appears to us that TD with pebble accretion has a significant
potential, and it must be studied further.

\begin{table*}
\caption{Selected simulations, their parameters and main results. The columns
  are: $M_p$, the initial planet mass, $\mj$; $f_{\rm p}$, the pebble mass
  fraction; $f_{\rm op}$, the opacity reduction factor; $M_{\rm d}$, the
  initial mass of the disc, in units of $\msun$; $\alpha_{SS}$, disc viscosity
  parameter; $M_{\rm end}$, mass of the planet in the end of the run, in units
  of $\mj$; $z_{\rm end}$, metallicity of the planet at the end of the
  simulation; $a_*$ [AU], the radial location of planet disruption or
  collapse.}
\begin{tabular}{lccccccccc}
Name & $M_p$ &   $f_{\rm p}$ & $f_{\rm op}$  & $M_{\rm d}$ &
$\alpha_{SS}$ & $M_{\rm end}$ & $M_{\rm core}$  & $z_{\rm end}$ & $a_{*}$ \\ \hline
M1LowOp1 &  1.0  &    0.0  &       1.0  &    0.12  &    0.01  &    0.00038  &
0.127  &    --  & 13.1 \\
M1LowOp4 &  1.0  &    0.0  &       0.1  &    0.12  &    0.01  &    0.00017  &
0.057  &    -- & 12.6  \\
M1LowOp7 &  1.0  &    0.0  &       0.01  &    0.12  &    0.01  &    1.0  &
0.045  &   0.015  &  61.1 \\ 
\hline
M5LowOp1 &  5.0  &    0.0  &       1.0  &    0.12  &    0.01  &    0.00017  &
0.18  &    --  & 6.76 \\
M5LowOp2 &  5.0  &    0.0  &       0.46  &    0.12  &    0.01  &    5.0  &
0.151  &    0.015  & 45.5\\
M5LowOp7 &  5.0  &    0.0  &       0.01  &    0.12  &    0.01  &    5.0  & 0.0
&    0.015 & 96.0  \\
\hline
M1Peb1 &  1.0  &    0.5  &       1.0  &    0.12  &    0.01  &    1.16  & 2.02
&    0.154   & 48.7\\
M1Peb3 &  1.0  &    0.158  &       1.0  &    0.12  &    0.01  &    1.16  & 3.65
&    0.152   & 2.83\\
M1Peb4 &  1.0  &    0.089  &       1.0  &    0.12  &    0.01  &    0.0025  & 0.84
&    --   & 3.97\\
\label{tab:1}
\end{tabular}
\end{table*}

\section{Conclusions}

In this article, steps are taken towards a numerically efficient way of
computing the planet-disc coevolution in the framework of the Tidal Downsizing
hypothesis for planet formation. Two ways of avoiding tidal disruption during
planet migration from the outer $\sim 100$~AU disc into the inner one are
considered: low dust opacity in the fragments and pebble accretion on to the
fragments. The former pathway requires extreme dust opacity reduction (by
$\sim 3$ orders of magnitude), predicts a negative correlation of giant planet
frequency of occurrence with metallicity, and leaves low mass $M_{\rm
  core}\sim 0.1 \mearth$ solid cores as disruption remnants. The significant
over-abundance of observed giant planets in metals \citep{MillerFortney11} is
also not addressed by this class of models. We conclude that fragments cooling
and contracting radiatively rather than due to pebble accretion are not likely
to explain observed planets. The most probable physical reason for this
irrelevance of low opacity models is that collisional grain fragmentation
keeps the supply levels of small grains sufficiently high \citep[as argued for
  protoplanetary discs by][]{DD05}, so that dust opacities never drop by
orders of magnitude below the interstellar values.

In contrast, high opacity ($f_{\rm op}\sim 1$) pebble accreting fragments
produce a positive metallicity correlation, yield massive solid cores as
remnants and also explain why giant gas planets are strongly over-abundant in
metals. On the basis of this paper, pebble accretion appears a key ingredient
to a successful TD model for planet formation. We hope this paper will serve
as impetus for the broader planet formation community to start investigating
TD in greater depth than has been achieved so far.

\section*{Acknowledgments}

Theoretical astrophysics research in Leicester is supported by an STFC
grant. The author thanks Richard Alexander, discussions with whom were very
influential in shaping the disc evolution model of the paper.  This paper used
the DiRAC Complexity system, operated by the University of Leicester, which
forms part of the STFC DiRAC HPC Facility (www.dirac.ac.uk). This equipment is
funded by a BIS National E-Infrastructure capital grant ST/K000373/1 and DiRAC
Operations grant ST/K0003259/1. DiRAC is part of the UK National
E-Infrastructure.

\label{lastpage}

\end{document}